\documentclass[aps,prb,twocolumn,superscriptaddress,amsmath,amssymb,showpacs]{revtex4}

\usepackage{graphicx}

\newcommand \be{\begin{eqnarray}}
\newcommand \ee{\end{eqnarray}}
\newcommand \ba{\begin{align}}
\newcommand \eea{\end{align}}

\begin{document}


\title{Instability types at ion-assisted alloy deposition: from two-dimensional to three-dimensional nanopattern growth}


\author{Gintautas Abrasonis}
\thanks{Corresponding author}
\email[]{g.abrasonis@hzdr.de}
\homepage[]{http://www.hzdr.de}
\affiliation{Institute of Ion Beam Physics and Materials Research, Helmholtz-Zentrum Dresden-Rossendorf, P.O.Box 51 01 19, 01314 Dresden, Germany}

\author{Klaus Morawetz}
\affiliation{M\"unster University of Applied Sciences,
Stegerwaldstrasse 39, 48565 Steinfurt, Germany}
\affiliation{International Institute of Physics (IIP),
Av. Odilon Gomes de Lima 1722, 59078-400 Natal, Brazil}
\affiliation{Max-Planck-Institute for the Physics of Complex Systems, 01187 Dresden, Germany}

\date{\today}

\begin{abstract}
Ion irradiation during film growth has a strong impact on structural properties. Linear stability analysis is employed to study surface instabilities during ion-assisted growth of binary alloys. An interplay between curvature-dependent ion-driven and deposition-driven  instabilities is investigated. We demonstrate that ion irradiation of growing binary alloys leads to the formation of composition-modulated surface patterns. It is shown that the ion-to-atom arrival ratio $R$ is the pattern control parameter. Close to the instability threshold we identify different regimes of instabilities driven by ion- or deposition-induced surface roughness processes, or roughness-composition feedback interactions. In particular, the synergistic effects of the curvature-dependent displacement and deposition coupling to the preferential sputtering or to the preferential diffusivity are found to induce instabilities and pattern formation. Depending on the film growth and ion-irradiation conditions, the instabilities show stationary or oscillating behavior. The latter one is exclusively connected with ion irradiation.  The corresponding phase diagrams are presented in terms of experimentally accessible parameters. This shows an alternative way to control surface patterning and to grow three-dimensional laterally or vertically ordered nanostructures.
\end{abstract}

\pacs{
81.16.Rf, 
79.20.Rf, 
81.15.Jj, 
81.15.Aa, 
68.35.Dv 
}

\maketitle


\section{Introduction}
Phase separation occurring during thin film growth results in a remarkably large variety of morphologies.
\cite{Venezuela:Nature1999,Petrov:JVSTA2003,Yasui:AM2007,Babonneau:PRB2005,Gerhards:PRB2004,He:PRL2006,Abrasonis:JAP2010,Abrasonis:APL2010,MacManus:NM2008} It takes place under 'frozen' bulk conditions \cite{Atzmon:JAP1992} where the bulk diffusion is restricted due to low growth temperatures. Such kinetic constraints for phase separation processes provide unique opportunities to influence the resulting structure down to the nanoscale. Interface and size effects of nanostructured materials synergistically act to influence the properties on the macro-scale. For phase separated, or nanocomposite films this leads to new properties and advanced (multi)functionality which cannot be predicted from the film constituents alone.  \cite{Ajayan:2005Book,MacManus:NM2008,Bonanni:PRL2008,Jamet:NM2006,Kuroda:NM2007,Mohaddes:NM2004,Zheng:Science2004,Fukutani:AM2004,Voevodin:JAP1997} Therefore, the control over the film nanostructure is of utmost importance.

Thin film growth includes many interacting kinetic processes (surface and bulk diffusion, repeated nucleation, shadowing, surface reaction, growth rate, etc.) which are influenced by external parameters such as temperature, growth rate, presence of energetic species, etc. \cite{Ohring:ThinFilms2006,Venables:2000Book,Thornton:JVST1986,Petrov:JVSTA2003} Structure zone models have been constructed to take into account the effects of temperature and thin film deposition assisting energetic ion bombardment. \cite{Movchan:1969,Messier:JVSTA1984,Thornton:JVST1986,Petrov:JVSTA2003,Anders:TSF2010} These models predict what characteristic structures will grow such as columnar grains, V-shaped columnar grains or equi-axed grains, etc, in certain experimental parameter ranges of thin film growth. Of particular importance are the surface instabilities because they can lead to composition nanopattern formation. \cite{Politi:PR2000} During the growth, such nanopattern is constantly buried by randomly depositing species. These species are again restructured due to the instability. In such a way the 2-dimensional (2D) structure appearing on the surface is transferred into the bulk, resulting in the formation of a 3-dimensional (3D) ordered heterogeneous structure. Considering the above, such a 3D nanostructure offers opportunities to control macro-scale properties. Therefore, the possibility to tune such surface instabilities is one of the paths to control the structure of the resulting films. The film  growth might give rise to one of the following types of instabilities \cite{Politi:PR2000}: 
\begin{itemize}
 \item dynamic instabilities due to interaction of impinging species with the film surface;  \cite{Raible:PRE2000,Evans:SSR2006} 
 \item kinetic instabilities if the growth of the surface is faster than its equilibration;   \cite{Politi:PR2000}
 \item thermodynamic instabilities where the desired structure is thermodynamically unstable;  \cite{Politi:PR2000}
 \item geometric instabilities where the surface roughness shadows the growth of the material. \cite{Politi:PR2000}
\end{itemize}
The analysis of the decomposition kinetics of phase-separating binary systems during the film growth has demonstrated that the resulting characteristic periodicity depends on the interplay of surface diffusivity and film growth rate. \cite{Atzmon:JAP1992,Tersoff:PRB1997,Leonard:PRB1997b,Yasui:AM2007,Fukutani:JJAP2008} If the growth rate surpasses a critical value any phase separation is suppressed. \cite{Tersoff:PRB1997,Leonard:PRB1997b} By performing linear and non-linear stability analysis, it has been shown that in the case of spontaneous phase separation surface roughness can couple to composition.
\cite{Leonard:PRB1997b,Leonard:PRB1997b,Leonard:PRB1997c} This phase separation can generate an elastic field which acts back on the film morphology. \cite{Leonard:PRB1997c} On the other hand, substrate mismatch and compositionally generated stress can also induce decomposition in otherwise thermodynamically stable alloys. \cite{Guyer:PRL1995,Guyer:PRB1996,Guyer:JCG1998,Tersoff:PRL1996,Leonard:PRB1998,Spencer:PRB2001} Kinetic effects such as differences in surface atomic mobilities can have a similar result. \cite{Venezuela:PRB1998}

In such a context, the possibility to use some external easily controllable factors whose influence and strength on the surface atomistic processes occurring during film growth would be comparable or even greater than the ones of 'intrinsic' processes is of particular importance. The energies delivered by ions to the (sub)surface atoms via atomic collisions far exceed the thermal budget of the atoms within the irradiated material. This creates unique conditions, which cannot be achieved by any other means. Ions provide an energy per material atom in the range of $\sim1-10$ eV/atom or more. That is much larger than the energy provided by any other processes such as solidification, recrystallization, interface energy minimization, thermal activation, phase transformation, etc. which are around or below $\sim1$ eV/atom. \cite{Harper:JVSTB1997}  Also the directionality is provided which can be externally controlled. \cite{Harper:JVSTB1997} Assisting ion irradiation is used to influence the structure of thin films. \cite{Nastasi:1996,Smidt:IMR1990} It is well established that ions have a huge impact on structural properties like density, grain size or texture. \cite{Nastasi:1996,Smidt:IMR1990} The control parameter is the ion-to-atom arrival ratio $R$. \cite{Nastasi:1996}  

It is well established that ion irradiation of surfaces induces pattern formation. \cite{Chan:JAP2007} For multicomponent materials such roughness pattern can couple to the composition resulting in compositionally modulated surface ripples \cite{Shenoy:PRL2007} or nanodots. \cite{Bradley:PRL2010} During ion erosion, the pattern formation depends on the composition \cite{Shenoy:PRL2007,Bradley:PRL2010} which is determined by the initial material composition or co-deposition rate of impurities. \cite{Bradley:PRB2011} Therefore, one can expect that such an ion irradiation also will affect the surface composition of the growing multicomponent thin film and will induce patterning of the growing surface. To our knowledge there are no theoretical studies of the ion effects on the surface compositional and spatial distribution during  film growth. Such an approach would present alternative ways to grow   
\begin{itemize}
 \item nanostructured surfaces with tunable surface roughness/composition patterns composed from a material different than that of the substrate; 
 \item 3D compositional nanopatterns or nanocomposites with tunable structural properties (periodicity, composition, tilt...).
\end{itemize}

In this paper, we employ the linear stability analysis on the surface roughness and composition during ion-assisted bi-component film growth. From the ion-induced effects we consider only those which are related to sputtering \cite{Bradley:JVSTA1988} and ballistic-induced surface redistribution fluxes. \cite{Carter:PRB1996,Davidovitch:PRB2007,Madi:PRL2011,Norris:NC2011,CCGGV11,CC12} Similarly we consider dynamic effects from the deposition. \cite{Raible:PRE2000,Evans:SSR2006} In order to highlight the effects of ion irradiation we consider only the case of an alloy deposition, i.e. there are no thermodynamic factors which could result in phase separation. \cite{Politi:PR2000,Spencer:PRB2001} 

The instability analysis is carried out in 'frozen bulk' approximation  where  majority of changes occur at the advancing film surface resulting in the 'frozen' structure in the bulk. \cite{Atzmon:JAP1992} In general, surface processes dominate in the temperature range $T< \sim 0.5 T_m$ ( $T$ - substrate temperature, $T_m$ - melting point).\cite{Thornton:JVST1986} Thus we focus on the growth situations where bulk atomic mobility is negligible in comparison to that of the surface. This is the main condition to transfer 2D patterning of the surface into the bulk. Otherwise, if the bulk evolves on itself, this patterning is lost. In this way the changes in the microstructure across the film cross-section represents the historical evolution of the microstructure of the film.\cite{Atzmon:JAP1992}

From the thermodynamic forces we will consider only those driven by surface concentration and curvature gradients, since these two are usually used in the literature concerning the pattern formation during ion irradiation. \cite{Chan:JAP2007,Shenoy:PRL2007,Bradley:PRB2011} The first counteracts the formation of composition differences while the second relaxes the surface  roughness modulations. 

We will not consider ad-atoms as separate surface species and no separate rate constant is introduced to describe the process of surface ad-atoms losing their mobility on the surface and their incorporation into the final location. \cite{Lichter:PRL1986} We will assume that surface atoms are available for diffusion as long as they are not covered by depositing fluxes. This is particularly valid for ion-assisted deposition as any atom can be knocked out of stable positions to become a mobile species on the surface. Therefore the characteristic time for diffusion (and any other induced process on the surface) is the interplay between the growth rate and surface diffusivity. \cite{Atzmon:JAP1992}

Throughout this paper we intend to explore the instabilities due to ion irradiation. The control parameters which are easily accessible from the experiments are the ion incidence angle $\theta$ and ion-to-atom arrival ratio $R$. It will be shown that these parameters drive the system from dynamic equilibrium to instabilities during the alloy growth. We will omit geometric effects due to deposition and consider the perpendicular depositing atom incidence which simplifies the analysis. On one hand the angular dependency of deposition is described well in the literature (see for example Refs. \onlinecite{Politi:PR2000,Lichter:PRL1986}). On the other hand we intend to separate these effects from the ion-beam induced ones. We will  demonstrate that ions induce composition-modulated surface-roughness patterns which show explicit dependencies on $R$. This originates directly from the form of continuum equations describing the ion-assisted alloy growth. The instabilities can be of stationary or oscillating type.

In the next section we develop the model of ion-assisted film growth which couples height and surface concentration of the binary alloy. In section III, the equations are linearized and the solutions are discussed with respect to time-growing modes.  In section IV, different types of instabilities are identified. Their conditions are linked to certain combinations of experimental parameters which reduces appreciably the parameter number. Section V summarizes the different cases in a mathematical form and develops the corresponding phase diagrams. The experimental realizations of the pattern formation are then discussed. Finally, section VI summarizes the findings of the paper and presents the conclusions. 

\section{Model development}

\subsection{Evolution equations}
\begin{figure}[h]
\includegraphics[width=8cm]{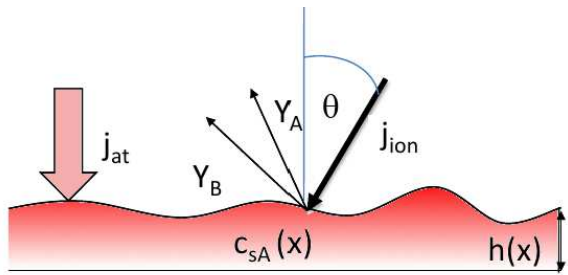}
\caption{\label{figschem1} The schematic picture of the experimental setup of depositing atomic fluxes $F_A$ and $F_B$ for two species under a bombarding ion flux $I$ creating a sputtering yield $Y_A$ and $Y_B$ for the two species.}
\end{figure}

Let us consider a surface in the $(x,y)$ plane with the local height $h(x,y)$. It exhibits the atomic density $\rho_S(x,y)$ and consists of the species A and B with the local surface atomic fraction $c_{S,i} (x,y)$, where $i=A,B$. 
The depositing flux $F=F_A+F_B$ of two species A and B (see Fig.~\ref{figschem1}) is assumed perpendicular to the film plane. The atomic surface ratio of species $i$ is $c_{0,i} = F_i / F$ and the arriving atoms of species $i$ stick to the surface with the probability $S_i$. Further we summarize the atomic flux 
\be
j_{\rm at}&=&\left (F_A S_A + F_B S_B\right )/\rho_S\approx \left ( c_{0,A}
  +c_{0,B}\right ) F S /\rho_S\nonumber\\
&=&F S /\rho_S 
\label{jat}
\ee 
where $c_{0,B}=1-c_{0,A}$ and we
assume that the mean sticking coefficient is independent of species $S\approx S_A \approx S_B$ and of local surface concentration. From now on we denote the fraction of the species $A$ in the incoming flux as $c_0$. 

The growing film surface is bombarded with ions. Ion beam irradiates the surface with the ion flux $I$ at the angle  $\theta$ to the surface normal. The projection $I {\rm cos}(\theta)$  of the ion flux density vector onto the film plane is along the $x$ axis. Similarly to the deposition flux ( Eq. (\ref{jat})) we normalize the ion flux  per surface atom, i.e. $j_{\rm ion} = I {\rm cos}(\theta) / \rho_S$. Irradiating ions induce sputtering of the species $i$ characterized by a sputtering coefficients $Y_i$, i.e. number of sputtered atoms of type $i$ per ion. The actual sputtering yield of the species $i$ with the surface fraction  $c_{S,i}$ is $Y_i c_{S,i}$. The total sputtering yield $\bar Y$ is the sum of the partial sputtering rates $\bar Y=Y_A c_{S,A} + Y_B c_{S,B}$.  

The evolution of the film surface height $h(x,y)$  in time $t$ follows from the conservation of matter at the solid surface \cite{Spencer:PRB2001,Pranevicius:SCT1995, Bradley:PRB2011}. Therefore let us consider an infinitely small surface area $\Sigma$ with the height $h(x,y)$. In the following we consider only linear analysis. Therefore, we neglect the difference between the local height normal to the surface and the projection on the $z$ axis, since this difference is quadratic in $\nabla h(x,y)$. \cite{Raible:PRE2000,Davidovitch:PRB2007} The time change of this material volume $\dot h(x,y) \Sigma$ is considered here due to the adsorption of atoms by atomic flux and the sputtering by ions as described above and the redistribution due to surface fluxes. Note that in contrast to the derivation presented in Ref. \onlinecite{Spencer:PRB2001} we do not assume a speed of the moving surface but obtain the speed from the competition of deposition, sputtering and relocation fluxes.

In the 'frozen' bulk approximation, all the processes take place within the 'active' volume $\Omega=\Delta \Sigma$ given by the monolayer thickness or atomic diameter $\Delta$ beneath the area $\Sigma$. Therefore the material volume is increasing with the atomic adsorption by the number of atoms per second $\sim \Omega j_{\rm at}$ with Eq. (\ref{jat}) and decreasing with the ion flux and sputtering yield $\sim -\Omega j_{\rm ion} \bar Y$. The rearrangement due to surface currents $\bf j_S$ is given by the change of the number of atoms  $N$ in the active volume $\sim \Omega \dot N$. The local fluctuating atomic density $n=N/\Omega$ changes as   $\dot n=-{\bf \nabla}\cdot {\bf j_S}/\Sigma$ where the current conservation was used. Similarly to the case of ion erosion with co-deposition of impurities \cite{Bradley:PRB2011}, collecting these contributions and dividing by the surface area $\Sigma$ with the help of Eq. (\ref{jat}) one gets 
\begin{eqnarray}
\frac{\partial h} {\partial t} &=& 
j_{\rm at}\Delta-j_{\rm ion}\Delta \bar Y- {\Delta^2} {\bf \nabla \cdot j_S}
\nonumber\\
&=&{\Delta  \over \rho_S} F S
- {\Delta  }\left ( {I\cos(\theta)\over \rho_S} \bar Y + \Delta {\bf \nabla \cdot j_S} \right ).
\label{height-conservation}
\end{eqnarray}

The current ${\bf{j_S}}={\bf (j_A+j_B)}$ describes the redistribution of the atoms on the surface due to various atomistic processes. These can be fluxes induced by deposition, ion irradiation or surface diffusion and are described in more detail below. Such fluxes change the local surface atomic fraction $c_{S,A}$ ($c_{S,B}$) of the element A (B). Multiplying the mass conservation for the species A $\dot n_A=-{\bf \nabla \cdot j_A}/ \Sigma$ with the active volume $\Delta \Sigma$ one obtains the contribution $\dot c_{S,A}\sim-\Delta {\bf \nabla\cdot j_A}$. Since this process originates from the active volume we have to make sure that the space was available for this process which means we have to multiply with the concentration of empty states $(1-c_{b,A})$ where $c_{b,A}=c_{S,A} (x,y,h(x,y) - \Delta)$  is the atomic fraction of the element A at the underlying layer. In the same way the effect of sputtering is considered as a kinetic process which decreases the concentration $c_{S,A}\sim -\!(1\!-\!c_{b,A}) c_{S,A}j_{\rm ion} Y_A $ and increases it by the complimentary process of species B, i.e.  $c_{S,A}\sim c_{b,A} (1-c_{S,A})j_{\rm ion} Y_B$. 

Similarly to Refs. \onlinecite{Pranevicius:SCT1995} and \onlinecite{Bradley:PRB2011}, by collecting these terms together the mass conservation law for the surface atomic fraction $c_{S,A}(x,y)$ of type A atoms results into
\ba
{\partial c_{S,A} \over \partial t} &= j_{\rm at}(c_0\!-\!c_{S,A})\!-\!(1\!-\!c_{b,A}) \left (c_{S,A}j_{\rm ion} Y_A \!+\!{\Delta}{\bf \nabla \cdot j_{A} }\right )\nonumber\\&
+c_{b,A} \left [(1-c_{S,A})j_{\rm ion} Y_B +{\Delta}{\bf \nabla \cdot j_{B} }\right ].
\label{concentration-conservation}
\end{align}
Please, note that the first term represents the source due to atomic deposition. \cite{Atzmon:JAP1992} The deposition of the atoms of the type B (A) on the atoms of the type A (B) reduces the local surface atomic fraction $c_{S,A}$ ($c_{S,B}$) of the element A (B), as the surface atoms of the type A (B) become the part of the bulk. The surface composition does not change if the same type of species are covered by the atoms from the vapor phase. On the other hand, sputtering or surface atomic redistribution removes the atoms of the type B (A) at the surface location $(x,y)$. The atoms from the underlying layer $h(x,y) - \Delta$ become atoms of the surface. The local surface composition changes only if the uncovered atoms from the bulk are of the type A (B). More detailed derivation of this balance can be found in the literature. \cite{Pranevicius:SCT1995,Bradley:PRB2011,Shenoy:PRL2007} Note, that in the presence of sputtering (with or without co-deposition) an altered layer forms close to the surface which differs in composition from that of the bulk. \cite{Pranevicius:SCT1995,Bradley:PRB2011} Therefore, the consideration of the altered near-surface composition depth profile in relation to the bulk concentration is necessary in deriving the correct concentration balance. \cite{Pranevicius:SCT1995,Bradley:PRB2011} Only in the particular case when the deposition rate is larger than the re-sputtering rate, the concentration distribution as a function of depth becomes homogeneous. \cite{Pranevicius:SCT1995} This allows a considerable simplification of the equations and will be considered below.

The first term on the right of Eq. (\ref{concentration-conservation}) describes the local composition changes due to deposition from the vapor phase. The second term describe the decrease rate of the atoms A  due to sputtering and surface redistribution as well as deposition fluxes of the atoms A, respectively. The third term describe the increase rate of the atoms A due to uncovering from the underlying layer at the depth $h(x,y) - \Delta$ resulting from the sputtering and surface redistribution as well as deposition of surface atoms of the type B, respectively.

\subsection{Planar film growth}

For the planar homogeneous film growth and absent surface redistribution fluxes ${\bf j_S}$, the surface height $h$ and atomic fraction $c_{A}$ are independent of $(x,y)$. The balance equations Eq. (\ref{height-conservation}) and Eq. (\ref{concentration-conservation}) result in the following system of equations
\ba
&{\partial h^{\rm planar} \over  \partial t} = j_{\rm at}\Delta-j_{\rm ion}\Delta\left[c_{S,A} Y_A +(1-c_{S,A}) Y_B\right]=V \nonumber\\
&{\partial c_{S,A} \over \partial t} = j_{\rm at}(c^0-c_{S,A})\nonumber\\&
\qquad \,\,\,\, -j_{\rm ion}\left[ c_{S,A}(1-c_{b,A}) Y_A -(1-c_{S,A})c_{b,A} Y_B \right]
\label{balance}
\end{align} 
where the coefficients denote the unperturbed, i.e. flat, film growth mode and $ V$ is the film growth rate. Remember that we have abbreviated for the atomic flux $j_{\rm at}=(F_A+F_B) S/\rho_S$, 
the projected ion flux $j_{\rm ion}=I {\rm cos} (\theta) / \rho_S$, $c_{b,A}=c_{S,A}(h-\Delta)$ is the concentration in the underlying layer or 'bulk'.

For steady state conditions there is a growth of a homogeneous film, i.e.
$h=h_0+ V t$. The element distribution becomes homogeneous over the depth, \cite{Pranevicius:SCT1995} i.e. 
$c_{A}^0 (h-\Delta)=c_{S,A}^0.$ 
Then the second equation of Eq. (\ref{balance}) yields
\begin{equation}
 c^{0}_{S,A}= {1\over 2RY}\left [1+R Y - \sqrt{1+R Y^2 -4c_0 R Y}\right ]  
 \label{surf-steady} 
\end{equation}
where $Y = Y_A-Y_B>0$ describes the preferential sputtering of one specie compared to the other one, where without loss of generality we assume $Y_A>Y_B$. The ion-to-atom arrival ratio $R=j_{\rm ion}/j_{\rm at}=(F_AS_A+F_B S_B)/I {\rm cos} (\theta)$ is already visible as the crucial ordering parameter.  This stationary solution is valid only for $ V>0$, i.e. when the growth rate is larger than the re-sputtering rate. From now on, we will denote that the surface concentration refers to the element $A$, i.e. $c^{0}_{S}=c^{0}_{S,A}$.

\subsection{Deposition-induced surface redistribution fluxes}

The surface height $h(x,y)$ evolution due to deposition in Eq. (\ref{height-conservation}) can be described by currents $\bf j^{\rm dep}$ representing
the dynamic interactions of incoming film forming species with the film surface and non-thermal particle rearrangements \cite{Kang:SS1992:269,Kang:SS1992:271} after the impact. \cite{Kang:SS1992:271} Therefore the deposition fluxes $F_i$, the sticking coefficient $S$, the surface density $\rho_S$ and the current terms on the right hand side of Eq. (\ref{height-conservation}) can depend on the surface slope and the curvature. \cite{Raible:PRE2000,Evans:SSR2006} Let us develop their contributions up to second order derivatives of the surface height.

Particles in the beam moving perpendicular to the substrate can be deflected when they come very close to the surface. This is because the atom-substrate inter-atomic forces act perpendicular to the surface, not the substrate. \cite{Raible:PRE2000} More atoms arrive at the surface places with ${\bf \nabla}^2 h < 0$ that at ${\bf \nabla}^2 h > 0$. This results into the so-called 'surface steering' effect  \cite{vanDijken:PRL1999} meaning that the deposition flux is not homogeneously distributed over the surface. \cite{vanDijken:PRL1999} The effect is more pronounced for lower atom kinetic energies and grazing incidences, but can also be active even for normal incidences. \cite{Montaleni:PRB2001}  Leaving only linear lowest-order terms, the deposition flux $F_i$ for the species $i$ can be expanded \cite{Raible:PRE2000}
\begin{equation}
  F_i= F_i^0 \left( 1 + a_{ST,i} {\bf \nabla}^2 h(x,y)  + ... \right)
\label{Steering}  
\end{equation}
where the constant $a_{ST,i}<0$ describes the dependence of the deposition flux $F_i$ on the surface curvature. Such a dependence results in a larger depositing atom flux at the crests than at the depressions. This induces a topographic instability.

The sticking coefficient $S$ can also depend on the local curvature. For example, for fcc(100) or bcc(100) geometries with four-fold hollow (4fh) adsorption sites, the density of 4fh \cite{Evans:SSR2006} is expected to be higher at local minima ${\bf \nabla}^2 h > 0$ than at local maxima ${\bf \nabla}^2 h < 0$. \cite{Evans:SSR2006} One can argue that this effect is not significant for amorphized surfaces, as they can easily relax to a random structure independent of the slope and curvature. Considering only linear lowest-order terms, this yields\cite{Evans:SSR2006} 
\begin{equation}
  S= S^0\left( 1 + a_{S} {\bf \nabla}^2 h(x,y)  + ... \right)
\label{Sticking}  
\end{equation}
where $a_{S}>0$ describes the dependence of the sticking coefficient $S$ on the surface curvature. Such a dependence of the sticking coefficient on the surface curvature has a stabilizing effect on the surface topography. 

If the adsorbing atoms do not hit directly an adsorption site, they still posses an ad-transient mobility with the significant component in the direction of the initial impingement and are 'funneled' downwards until they reach such an adsorption site. \cite{Evans:PRB1991} Downward funneling is a non-thermal process. It induces a downward flux also described by the second term on the right hand side of Eq. (\ref{height-conservation}). Leaving the linear lowest-order terms one obtains  \cite{Kang:SS1992:271}
\begin{equation}
  {\bf j}^{\rm ne}_i =  - \frac {F_i^0 S^0}{\rho_{S}^0} a_{DF,i} {\bf \nabla} h(x,y)  +...
  \label{Downward-funneling} 
\end{equation}
where $a_{DF,i}>0$ describes the mobility of funneled atoms and $\rho_{S}^0$ describes the atomic density of the defect free surface. This effect has again a stabilizing behavior on the surface topography.

Deposited atoms can be trapped on the sides of the slopes instead of funneling down to adsorption sites. \cite{Caspersen:PRB2001} Such 'restricted downward funneling' can be represented by expanding $\rho_S$ \cite{Caspersen:PRB2001}
\begin{equation}
{1\over\rho_S}={1\over \rho_{S}^0} \left (1 + a_{SD}{\bf \nabla}^2 h(x,y) +...\right ).
\label{surface-density}
\end{equation}
The term $a_{SD}<0$ describes the defects which trap the funneling atoms and which are dependent on the curvature. It has a destabilizing effect on the surface topography

Combining Eqs. (\ref{Steering}), (\ref{Sticking}), (\ref{Downward-funneling}), (\ref{surface-density}) into Eq. (\ref{height-conservation}) and leaving only the linear order terms, one gets
\ba
{\partial h /  \partial t}  \approx j_{\rm at}\Delta&-j_{\rm ion} \Delta\left[c_{S,A} Y_A +(1-c_{S,A}) Y_B\right] 
\nonumber\\
&-\Delta^2{\bf  (\nabla\cdot j_{A}^{\rm dep} + \nabla\cdot j_{B}^{\rm dep})}
\label{final-deposition}
\end{align}
with
\be
{\bf j}_{i}^{\rm dep} &=& - {F_i^0 S^0 \over \rho_{S,0}}\left (a_{ST,i} + a_{S} + a_{DF,i} + a_{SD}\right ) {\bf \nabla} h 
\nonumber\\
&=& -c_{0,i}{j_{\rm at}}S^{\rm dep}_i {\bf \nabla} h  
\label{deposition-instabilities}
\ee
where $S^{\rm dep}_i$ combines the dynamic surface slope and curvature effects of species $i=A,B$. Its sign depends on the relative magnitude of each of the processes and can have either positive (stabilizing) or negative (destabilizing) effect. For a given material system, there are only few opportunities to control these dynamic effects including the incidence angle. Since we assume the depositing flux incidence angle fixed, $S^{\rm dep}$ will be considered as a constant further on. In the following, we will call the parameter $S^{\rm dep}$ the curvature-dependent deposition coefficient.

\subsection{Ion irradiation - induced fluxes}

When penetrating into the solid, energetic ions  create recoils which retain to the large extent the initial momentum of the incoming ions before being stopped. \cite{Carter:PRB1996,Davidovitch:PRB2007} If the ion beam is oblique to the substrate, then there is a component along the sample surface. Due to the incompressibility of the solid, the excess of the material density relaxes into the surface. \cite{Carter:PRB1996,Davidovitch:PRB2007} For the ion flux $I$ with the incidence angle $\theta$ from the surface normal, this generates ion-induced ballistic  relocation currents proportional to $I {\rm cos} (\theta) {\rm sin} (\theta)$, where the product with ${\rm cos} (\theta)$ reflects the projection of the ion beam on the surface while the product with ${\rm sin} (\theta)$ reflects the projection of the recoil atom flux along the surface. Expansion to linear order yields the following expression for the ion-induced relocation currents \cite{Carter:PRB1996,Davidovitch:PRB2007,Madi:PRL2011,Norris:NC2011}
\begin{equation} 
 {\bf j}_i^{\rm rel}=-c_{S,i} j_{\rm ion} S^{\rm rel}_i \left \{  S^{\rm rel}_x(\theta) {\partial h \over \partial x} , S^{\rm rel}_y(\theta) {\partial h \over \partial y}   \right \} 
\label{reloc_coeff}
\end{equation} 
where $S^{\rm rel}_i=f_i d_i$  is the relocation yield of the species $i$, $f_i$ is the number of relocated atoms and $d_i$ is the average relocation distance. The dependence on the ion-incidence angle is explicitly included in $S^{\rm rel}_{x,y} (\theta)$. It can change the sign and thus can act to stabilize or destabilize the surface roughness. \cite{Davidovitch:PRB2007,Madi:PRL2011,Norris:NC2011} Assuming a simple angular dependence $\sim {\rm cos} ( 2 \theta)$ from the model of Carter and Vyshniakov,  \cite{Carter:PRB1996} one can see that $S^{\rm rel}_{x,y} (\theta)$ becomes negative for $\theta>45^{\circ}$. This means that the areas with positive curvature ($\partial^2 h / \partial x,y^2 > 0$) or depressions start to loose the material. The latter starts to accumulate at the areas with negative curvature $\partial^2 h / \partial x,y^2 < 0$) or crests, i.e. the instability grows. For silicon, it has been demonstrated that  such ion-induced relocation currents have the dominating effect for the ion-induced surface topographic instabilities. \cite{Madi:PRL2011,Norris:NC2011} 

Part of the ion-displaced atoms gain enough energy and leave the solid, i.e. are sputtered. The sputtering can depend on the local surface curvature. \cite{Bradley:JVSTA1988} To first-order this yields \cite{Shenoy:PRL2007,Bradley:JVSTA1988}
\ba
Y_i=Y^0_i\left[ 1+ \upsilon_0 {\partial h \over \partial x} - S^{\rm sp}_x (\theta) \Delta {\partial^2 h \over \partial x^2}- 
 S^{\rm sp}_y (\theta) \Delta {\partial^2 h \over \partial y^2}\right ]. 
\label{sputter-coeff}
\end{align}
If the sign of either $S^{\rm sp}_{x,y}(\theta)$  becomes negative for a certain angular range of $\theta$, it results in larger sputtering yields at the depressions than on the crests, i.e. the surface roughness becomes unstable. \cite{Bradley:JVSTA1988} While for Si this effect has been shown to be weak, \cite{Madi:PRL2011,Norris:NC2011} the relative strength of the curvature-dependent sputtering in relation to the ion-induced relocation remains to be determined for other material systems. Therefore, for the sake of completeness it will be considered in the further analysis.

\subsection{Diffusive currents}

Surface diffusion fluxes are proportional to the gradient of the surface chemical potential $\mu_S$ and the surface atom diffusivity $D_i$ \cite{Spencer:PRB2001}
\begin{equation}
{\bf j_i}^{\rm diff} = {\Delta \rho_S D_i \over k_BT} {\bf \nabla} \mu_S
\label{miu-fluxes}
\end{equation}
Considering only the effects of concentration gradients and surface 'stiffness' or capillarity, this leads to \cite{Spencer:PRB2001,Shenoy:PRL2007}
\begin{equation}
	{\bf j_i}^{\rm diff} = -\Delta D_i \rho_S {\bf \nabla} c_{S,i} + c_{S,i} {\Delta D_i \rho_S \Omega \gamma \over k_BT} {\bf \nabla} ({\bf \nabla})^2 h
	\label{diffusion-fluxes}
\end{equation}
where $\Omega$, $\gamma$, $T$, and $k_B$  stand for the atomic volume, surface energy, temperature and Boltzmann's constant, respectively.

\subsection{Final set of equations}

Taking into account Eqs. (\ref{final-deposition})-(\ref{diffusion-fluxes}), mass conservation (Eqs. (\ref{height-conservation}) and (\ref{concentration-conservation})) leads to  the following set 
\ba
&{\partial h \over \partial t} =j_{\rm at} \Delta - j_{\rm ion} \Delta \left [ c_{S,A} Y_A  + (1-c_{S,A}) Y_B  \right] 
\nonumber\\
&- \Delta^2 {\bf \nabla}\cdot\left \{{\bf j_A}+{\bf j_B}\right \}  \nonumber\\
&{\partial c_{S,A} \over  \partial t} = j_{\rm at} (c^0\!-\!c_{S,A}) \!-\! (1\!-\!c_{S}^0) \left[j_{\rm ion}c_{S,A} Y_A  + \Delta {\bf \nabla} \cdot {\bf j_A}  \right] 
\nonumber\\&
+c_{S}^0 \left[ j_{\rm ion}(1-c_{S,A}) Y_B  + \Delta {\bf \nabla} \cdot {\bf j_B}  \right] 
\label{mass-conservation}   
\end{align}
where ${\bf j_i}={\bf j_i}^{\rm diff}+{\bf j_i}^{\rm rel}+{\bf j_i}^{\rm dep}$  is the sum of the diffusive flux, Eq. (\ref{diffusion-fluxes}), relocation, Eq. (\ref{reloc_coeff}), and deposition Eq. (\ref{deposition-instabilities}). In deriving Eq. (\ref{mass-conservation}) we used again the fact that for film-growing conditions the bulk concentration is equal to the steady-state surface value $c_{b,A}=c_{S}^0$. \cite{Pranevicius:SCT1995} 

\begin{table*}
\caption{\label{table1}. Coefficients of the terms of the right-hand side of Eq. (\ref{linear-equations}). Note that the factor $\rho_s /\Delta/j_{\rm at}$ can be absorbed into dimensionless diffusivities which would result into a single coefficient in $B$ and $B^*$ of $\omega=\gamma \Delta^2/T$.}
\begin{align*}
&\begin{array}{llllll}
Y&=&(Y^0_A-Y^0_B)& 
Y^*&=&(1-c^0_{S})Y^0_A + c^0_{S} Y^0_B \\
\upsilon &=& \upsilon^0 [c_{S}^0 Y^0_A+ (1-c_{S}^0) Y^0_B]  & 
\upsilon^* &=& c^0_{S} (1-c^0_{S}) \upsilon^0 (Y^0_A - Y^0_B) \\
D&=&\left( D_A - D_B \right )\rho^0_S /j_{\rm at} & 
D^*&=& \left[ (1-c^0_{S}) D_A + c^0_{S} D_B \right ] \rho^0_S /j_{\rm at}\\
B &=& \left[   c^0_{S} D_A +( 1-c^0_{S}) D_B \right ] \rho^0_S \Delta^2 \gamma / k_B T j_{\rm at}& 
B^*&=& c^0_{S} (1-c^0_{S}) \left( D_A - D_B \right ) \rho^0_S \Delta^2 \gamma / k_B T j_{\rm at}  \end{array}\\\\
&\begin{array}{lll}
S^{\rm ion}_{x,y}&=& S^{\rm sp}_{x,y}(\theta) \left[ c^0_{S} Y^0_A  + (1-c^0_{S}) Y^0_B  \right] +  S^{\rm rel}_{x,y}(\theta) \left[ c^0_{S} S^{\rm rel}_A  + (1-c^0_{S}) S^{\rm rel}_B  \right] \\
S^{\rm dep}_{x,y}&=&  c_0 S^{\rm dep}_A  + (1-c^0) S^{\rm dep}_B  \\
S_{x,y}&=& S^{\rm ion}_{x,y}+S^{\rm dep}_{x,y}/R\\
 S^{*,ion}_{x,y}&=& c^0_{S}(1-c^0_{S})  \left [
S^{\rm sp}_{x,y}(\theta) \left (Y^0_A- Y^0_B \right ) + S^{\rm rel}_{x,y}(\theta) \left (S^{\rm rel}_A- S^{\rm rel}_B \right )\right ] \\
S^{*,\rm dep}_{x,y} &=& c_0(1-c^0_{S}) S^{\rm dep}_A - c^0_{S}(1-c_0)S^{\rm dep}_B = c_0(1-c^0_{S}) (S^{\rm dep}_A-S^{\rm dep}_A)+(c_0-c^0_{S})S^{\rm dep}_{x,y} \\
S^{*}_{x,y}&=& S^{*,\rm ion}_{x,y} + S^{*,\rm dep}_{x,y}/R \\
\end{array}
\end{align*}
\end{table*}

\section{Linear analysis}

\subsection{Linearized equation and parameters}

Now we are going to study the stability of this solution to perturbations $u(x,y)$ and $\phi(x,y)$ in height and surface concentration, respectively,
\begin{eqnarray}
&h=h_0+V t+u(x,y), \quad
&c_{S,A}=c^0_{S}+\phi(x,y).
\label{balance-solutions}
\end{eqnarray} 
Introducing Eq. (\ref{balance-solutions}) into Eq. (\ref{mass-conservation}), changing the variables $\tau=j_{\rm at} t$ and expressing the height and the spatial coordinates in units of the monolayer thickness ($U=u/\Delta$, $x\rightarrow \Delta
x$) we get the following system of equations
\begin{eqnarray}
{\partial U \over \partial \tau} &=& -  R \upsilon \frac{\partial U}{\partial x} + R S_x  \frac{\partial^2 U}{\partial x^2} + R S_y \frac{\partial^2 U}{\partial y^2} 
\nonumber\\&&
-R Y \phi + D \nabla^2 \phi -B \nabla^4U  \nonumber\\
{\partial \phi \over  \partial \tau} &=& -\phi - R Y^* \phi  -  R \upsilon^* \frac{\partial U}{\partial x} + R \left[S^*_x  \frac{\partial^2 U}{\partial x^2} + S^*_y  \frac{\partial^2 U}{\partial y^2} \right] 
\nonumber\\&&
+ D^* \nabla^2 \phi -B^* \nabla^4U 
 \label{linear-equations}   
\end{eqnarray}
with $R=j_{\rm ion}/j_{\rm at}$ and where the coefficients are summarized in Table \ref{table1}. The seemingly cumbersome combination of parameters bear a clear physical meaning. The $^*$ parameters denote the coefficients in the concentration equation of Eq. (\ref{linear-equations}) while all the others are related to the equation describing the surface roughness evolution.

The terms $\upsilon \partial U / \partial x$ and $\upsilon^* \partial U / \partial x$ in Eq. (\ref{linear-equations}) result in a surface drift. As the surface is constantly buried by incoming species, these drift terms determine the 3D-nanopattern tilt in relation to the sample surface. One can absorb the surface drift term $\sim \partial U / \partial x$ by a running frame transformation $x\to x-\tau R \nu$ in the surface roughness equation of Eq. (\ref{linear-equations}). However, the drift term in the concentration equation will induce nontrivial patterns. In the following we neglect the drift terms restricting to small drift velocities and devote this analysis to a later investigation.

To illustrate the usefulness of these parameters consider the diffusion terms $D, D^*$ and the capillarity $B, B*$ terms in Eq. (\ref{linear-equations}) only. While the $D, B^*$ terms can change sign due to the difference between species parameters and can therefore act to destabilize the system, the $D^*, B$ parameters are always positive due to sum of parameters, see Table \ref{table1}. 
In the absence of the deposition flux ($j_{\rm at}=0$) and ion-irradiation ($R=0$), one has always $D^*>D$ and $B>B^*$ and any infinitesimal composition or surface roughness fluctuation is suppressed.

\begin{figure}[h]
\includegraphics[width=8cm]{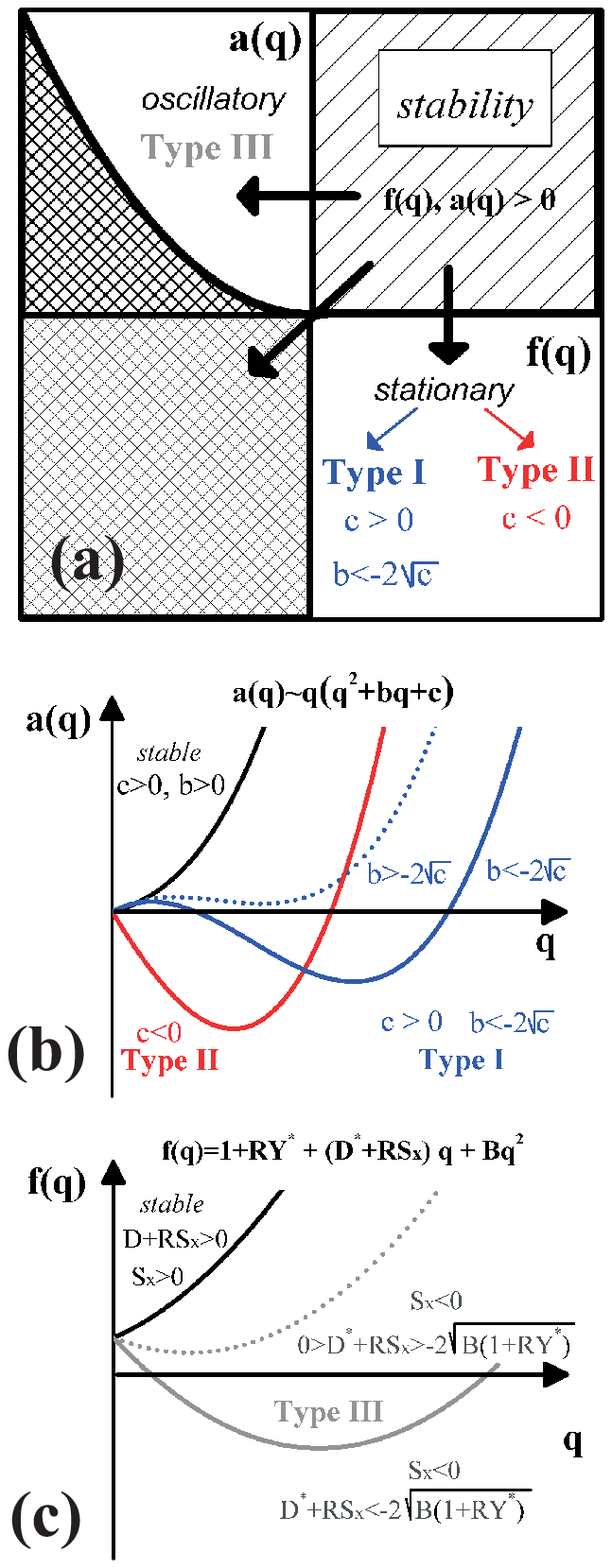}
\caption{\label{Fig2} (a) The different regimes of unstable modes in terms of (b) the parameter $a(q)$ and (c) the parameter $f(q)$ corresponding to Eq. (\ref{bc-coeff}). The arrows indicate possible reachable instability regions with continuous change of R.}
\end{figure}

In the following, the term 'displacement' will comprise both sputtering and relocations. \cite{Abrasonis:SCT2002} The driving terms for instabilities are $S_{x,y}$ and $S^*_{x,y}$ for surface roughness and composition, respectively. These terms describe the curvature-dependent total and preferential displacement/deposition rates, respectively.  These terms are multiplied by the surface curvature $\partial^2 U / \partial x^2$ (see Eq. (\ref{linear-equations})). Positive (negative) sign of $S_{x,y}$ means that there is a material loss (gain) at the crests ($\partial^2 U / \partial x^2 <0$) and a material gain (loss) at the depressions ($\partial^2 U / \partial x^2 <0$). Thus $S_{x,y}>0$ ($S_{x,y}<0$) translates into a stabilizing (destabilizing) action on the surface roughness. Otherwise, a positive(negative) sign of $S^*_{x,y}$ means that there is a loss (gain) at the crests ($\partial^2 U / \partial x^2 <0$) and gain (loss) at the depressions ($\partial^2 U / \partial x^2 <0$) of the preferentially displaced material. Thus this term acts in destabilizing the composition in all the cases except for the situation with no preferential displacement or deposition (both alloy components have similar ballistic properties) or with flat surfaces ($\partial^2 U / \partial x^2 =0$). Therefore surface-roughness instabilities can start independently from the surface composition effects while  surface composition instabilities cannot occur without coupling to the surface roughness. In conclusion, instabilities can occur via surface roughness effects or via composition-roughness feedback interactions.

The curvature-dependent deposition terms $S_{x,y}^{\rm dep}$ and $S_{x,y}^{*, \rm dep}$ are proportional to the incoming flux $j_{at}$ (Eqs. (\ref{Steering}) - (\ref{final-deposition})). In case that the deposition has a destabilizing effect, $j_{at}$ is a control parameter. The terms $\sim R S_{x,y}=R S_{x,y}^{\rm ion}+S_{x,y}^{\rm dep}$ and $\sim R S^*_{x,y}=R S_{x,y}^{*\rm ion}+S_{x,y}^{*\rm dep}$  explicitly have linear dependencies on $R$ which controls the degree of ion-induced effects compared to deposition effects. While the exact dependence is determined by the details of collision cascades, \cite{Davidovitch:PRB2007} we should note that both $S^{\rm sp}_{x,y} (\theta)$ and $S^{\rm rel}_{x,y} (\theta)$ can change their sign resulting in stabilizing or destabilizing behavior. This can be controlled by varying the ion-incidence angle. \cite{Davidovitch:PRB2007} In addition, the sign change of $S^{\rm rel}_x (\theta)$ and $S^{\rm sp}_x (\theta)$ is governed by different properties of ion induced collisions cascades and thus occur in different angular ranges. \cite{Davidovitch:PRB2007} As it will be shown in the next sections, such general properties of Eq. (\ref{linear-equations}) and the dependence of ion-induced effects on $R$ and $\theta$ create a rich variety of pattern formations besides those induced by deposition.

As the curvature-dependent terms are fixed, the ratio of ion to deposition beam $R$ acts as a control parameter. If the ion-induced curvature-dependent ballistic term combinations ($R\ne 0$ terms) become negative for a certain ion-incidence angle $\theta$ range, their degree of influence on the surface dynamics is 'regulated' by $R$: the higher $R$, the larger the influence of ion effects. If deposition dynamics leads to surface stabilization, then at some critical $R_{\rm crit}$ instability is expected to result in a growth of a pattern. In turn, if the deposition dynamics leads to surface destabilization, one can find a certain ion-incidence angle $\theta$ range, where the ion-induced curvature-dependent ballistic term combinations become positive. After reaching a critical ion-to-atom arrival ratio $R_{\rm crit}$ the growing surface becomes stabilized.

\subsection{Stability of modes}

Let us perform a linear stability analysis when the pattern growth occurs. For this we consider only a 1D case along $x$. The solution of Eq. (\ref{linear-equations}) in the form of 
$(U,\phi)=(U^0,\phi^0) \times exp \left( ik x + r t \right)$ represents the growth rate $r$ as eigenvalues of the matrix 
\begin{equation}
{\textbf A} = \left(
  \begin{array}{lr}
    -R S_x k^2 - B k^4     & -R Y - D  k^2 \\
    -R S^*_x k^2 - B^* k^4 & -1-R Y^* - D^* k^2 \\
  \end{array}
 \right)
\label{matrix}
\end{equation}
with the eigenvalues  
\be
  r&=&\frac{1}{2} \left ( -f(q) \pm \sqrt{ f(q)^2 - 4 a(q)}\right ),\,\,{\rm where}\nonumber\\
f(q)&=&-tr{\textbf A} = B \left( d+ e\, q + q^2   \right) \nonumber\\
a(q)&=&{\rm det} {\textbf A}= f_{\rm dis} \, q \left ( c + b\, q + q^2  \right ),
\label{eigenvalue}
\ee
$k^2=q\ge 0$ and the auxiliary quantities
\be
f_{\rm dis} &=& BD^* - B^*D={\rho_S^2 \gamma \Delta^2\over k_B T  j_{\rm at}^2}D_A D_B>0\nonumber\\
b&=&\frac{B + R(S_x D^* - S^*_x D + BY^* - B^*Y)}{f_{\rm dis}} \nonumber \\
c&=&\frac{R \left[ S_x  + R(S_x Y^* - Y S^*_x) \right]}{f_{\rm dis}}\nonumber\\
d&=&{1 + R Y^*\over B}>0\nonumber\\
e&=&{D^*  + RS_x \over B}. 
\label{bc-coeff}
\ee
For a given wave number $k$, a positive (negative) $r$ value indicates that this mode is unstable (stable) and will grow (be suppressed) in amplitude. The growth rate $r$ is negative (stable) if $a(q)>0$ which is fulfilled in the absence of deposition and ion irradiation since always $f_{\rm diss}>0$. The combination of parameters in $f_{\rm diss}$ acts as a 'dissipative' force bringing the system to and maintaining at the equilibrium when external influences are absent. 
In order to determine the band of unstable vectors $k$, we proceed first with the stability analysis \cite{Cross:2009} illustrated in Fig.~\ref{Fig2} . The system is stable for $a(q)>0$ and $f(q)>0$ since then $r<0$ for all $k$ and has oscillatory solutions if $f(q)^2<4 a(q)$. 
In the absence of ion irradiation and deposition, the condition $f(q)>0$ is always satisfied (see also the discussion above). Within the continuous change of the control parameters we can reach the three adjacent instability regions from the stable one by the three arrows indicated in Fig.~\ref{Fig2} (a). The left lower quarter in Fig.~\ref{Fig2}, corresponding to $f(q)<0$ and $a(q)<0$, can only be reached via a special point $a(q)=0$ and $f(q)=0$. This condition requires both $a(q)$ and $f(q)$ to become zero for the same value of $R$ and/or deposition $S^{\rm dep}$. This can be reached only with a unique combination of parameters and maximal 2 special wave numbers. This regime does not provide any control means and therefore is of low relevance to experiments. We are left with two different possible paths from stability to instability: (i) $a(q)<0$ and $f(q)>0$ for stationary- growing patterns and (ii) $f(q)<0$ and $a(q)> (f(q)/2)^2 >0$ for oscillatory patterns since the square-root term becomes purely imaginary for the growth rate $r$ in Eq. (\ref{eigenvalue}).

\subsection{Stationary patterns}

Stationary instabilities occur when $f(q)$ stays positive and $a(q)$ changes its sign to become negative (see Fig.~\ref{Fig2} (a))
. This occurs in the range of the two $q$ values
\be
q_0^\pm=\frac 1 2 \left (-b\pm \sqrt{b^2-4c}\right )
\label{solq}
\ee
determined by the sign of parameters $b$ and $c$ which is illustrated in Fig.~\ref{Fig2} (b). We see that two real $q$ occur for either  $c<0$  or  $c>0$, $b<-2 \sqrt{c}$.

Before continuing to discuss the different instability types of these two cases lets discuss the physical meaning of the terms $b$ and $c$ which are crucial for the type of instability. The term $b$ has three contributions  (Eq. (\ref{bc-coeff})):

\begin{itemize} 
\item An $R$-independent positive term $B/f_{\rm dis} \approx j_{\rm at} \times \gamma \left[ c^0_{S} D_A +( 1-c^0_{S}) D_B \right ] >0$ related to the film growth rate and capillary forces. This term stabilizes the surface composition and roughness. It increases concomitantly with the deposition flux $j_{\rm at}$. This term always renders $b>0$ for low values of $R$;
\item A term $R( BY^* - B^*Y)$ describing how preferential sputtering interacts with the curvature to destabilize the surface. This term is always positive as $Y^*>0$, $Y^*>Y$, $B>0$, $B>B^*$. This reflects the fact that both surface capillary forces $B$ and total sputtering rate $Y^*$ act in stabilizing the surface roughness and composition, see Eq. (\ref{linear-equations});
\item A term $R(S_x D^* - S^*_x D)$  describing how the curvature-dependent (total or preferential) displacement/deposition processes interact with diffusion to (de)stabilize the surface. Only this term can lead to $b<0$, thus to pattern formation.
\end{itemize}

The term $c$ in Eq. (\ref{bc-coeff}) has two contributions: $c \sim R \left[S_x + R(S_x Y^* - Y S^*_x) \right]$. The first one is $R S_x = R S^{\rm ion}_x + S^{\rm dep}_x$. This is the {\it total} curvature-dependent displacement/deposition coefficient. The second one represents (de)stabilization related to the interaction between curvature-dependent displacement/deposition and sputtering (both total and preferential). At $R \approx 0$ (weak or absent ion irradiation), the sign of the curvature-dependent deposition term $S^{\rm dep}_x$ determines the slope of $a(q)$ growing for small wave numbers $q$ towards negative or positive values, and thus determines the instability.

\subsection{Oscillatory patterns}

When ion irradiation or/and deposition is present, oscillatory behavior might appear if $f(q)<0$ or $e<-2 \sqrt{d}$. Such condition yields (see Eqs. (\ref{bc-coeff}))

\begin{equation}
R S_x<- \left (2 \sqrt{B(1+ R Y^*)} + D^* \right). 
\label{oscilatory-condition}
\end{equation}

As $R>0$ and the expression in the brackets on the right  side of Eq. (\ref{oscilatory-condition}) is positive, this necessarily means $S_x < 0$ and the total curvature-dependent displacement/deposition term must become destabilizing. In addition, this also means that the instability can occur only when the control parameter $R>0$ exceeds a certain critical value. 
Following the discussion above the sign change of $f(q)$ is driven by the interplay between the ion-irradiation/deposition induced surface instability and the surface diffusion and preferential sputtering rate. This is because the curvature-dependent displacement/deposition $S_x<0$ destabilizes while the surface diffusion $D^*>0$ and preferential sputtering $Y^*>0$ stabilize the system.

\section{Types of instabilities}

In order to understand the interplay of ion-irradiation and deposition effects to induce instabilities, we will concentrate separately on the effects produced by ion irradiation and deposition. Their combined effect will be discussed at the end. Therefore let us consider first a planar growth of the surface ($S^{\rm dep}=0$) and concentrate exclusively on the instabilities induced by curvature-dependent ion displacements. 

\subsection{Ion-irradiation induced instabilities}
\label{section-ion-instabilities}

\subsubsection{Instabilities driven by linear terms in $q$ (type II)}

\paragraph{Type IIa: curvature-dependent displacement-driven pattern formation.}

First we discuss the case when $c<0$. This condition can be satisfied for two possibilities $S_x\lessgtr 0$ according to Eq. (\ref{bc-coeff}). As seen in Fig. \ref{Fig2} (b), in both cases the instability starts from $q=0$ and  the width of the band  $\Delta q$ of unstable modes is equal to the positive solution $q_0^+$ of Eq. (\ref{solq}). For the case of $S_x<0$, expanding in powers of $R$ we get from Eq. (\ref{bc-coeff}) and (\ref{solq}) the bandwidth of unstable wave vectors
\begin{equation}
\Delta k^2 = \Delta q \approx -R \frac{S_x}{B}+o(R^2).
\label{deltak-1}
\end{equation}

Near the threshold where $a(q)$ changes the sign it becomes small so that $r\approx -a(q)/f(q)$. The maximal growth rate  is approximately given by the condition  $ d a(q) / dq =0$ which leads to
\begin{equation}
k_{\rm max}^2=q_{\rm max}=\frac 1 3(\sqrt{b^2-3 c}-b)  \approx -R\frac{S_x}{2 B}+o(R^2).
\label{kmax-1}
\end{equation}

The most unstable wavelength diverges when $R\rightarrow 0$ while the band width shrinks as $\sim\sqrt{R}$. This is the so-called type-II instability \cite{Cross:2009}. This instability is characteristic for ripple formation during ion erosion.
\cite{Norris:NC2011,Madi:PRL2008} Both $k_{\rm max}$ and $\Delta k$ depend only on the parameters related to the surface roughness. Therefore, in this regime ($c<0$), the pattern formation is driven by ion-induced surface-roughening processes. A surface roughness pattern develops first. The segregation follows due to the $h(x,y)$-$c_{S}(x,y)$ coupling.

Note, that the condition $S_x<0$ requires some further restrictions to render $c<0$. If $S_x Y^* - Y S^*_x<0$ the instability condition $c<0$ is fulfilled for all $R$ . If $S_x Y^* - Y S^*_x>0$ the ion-to-atom arrival ratio should not exceed a critical value $R<R_{\rm crit,1}$ with
\be
R_{\rm crit,1} = -{S_x \over S_x Y^* -S^*Y}.
\label{Rcrit1}
\ee

\paragraph{Type IIb: preferential curvature-dependent displacement and sputtering.}

For the case $c<0$, let us consider now  the possibility  $S_x>0$, i.e. the total curvature-dependent displacement acts in stabilizing the surface. Such conditions requires $S_x Y^* - Y S^*_x<0$ (see according to Eq. (\ref{bc-coeff}). As both $S_x$ and $Y^*$ (total sputtering coefficient) are positive, the instability is driven by the combination $YS^*_x>0$, i.e. by preferential sputtering and preferential curvature-dependent displacement. The condition $c<0$ requires that these both terms must have the same sign ($S^*_x>0$ as $Y^*>0$ by definition).

This represents the following scenario: 

\begin{itemize}
\item the element segregation occurs on the surface as the 
preferentially displaced element is depleted from the crests and enriched in the depressions (term $R S^*_x \frac{\partial^2 U}{\partial x^2}$ in Eq. (\ref{linear-equations}),
\item preferential sputtering of the same element occurs (term $-R Y \phi$ in Eq. (\ref{linear-equations}), 
\item as this element tends to segregate at the depressions, preferential sputtering of the depressions takes place which results in the increase of the surface roughness, i.e. an instability occurs.  
\end{itemize}

The instability is also of type II following the classification by Cross\&Greenside \cite{Cross:2009}. The main difference to the type II$_a$ instability is that the ion-to-atom arrival ratio $R$ must be larger than a critical value Eq. (\ref{Rcrit1}). Close to the instability threshold one can write $R=R_{\rm crit,1} + \delta$. Introducing this expansion into Eq. (\ref{solq}) and leaving only linear terms in $\delta$ to get $\Delta c=-{S_x/f_{\rm dis}}\delta $ and the following expression for the band width $\Delta q$ 
\begin{equation}
\Delta k^2 = \Delta q \approx {S_x\over f_{\rm dis} b_{\rm crit}}\delta 
\label{deltaq-II}
\end{equation}
where $b_{\rm crit}$ is the value of the parameter $b$ at $R_{\rm crit,1}$. The bandwidth shrinks again as $\Delta k \sim \sqrt{\delta}$. The maximal growth appears again as in Eq. (\ref{deltak-1}) and Eq. (\ref{kmax-1}) nearly at the minimum of $a(q)$ leading to $k_{\rm max}^2=q_{\rm max}\approx \frac { \Delta q}{2}$ as above.

Note, that the destabilizing effect is non-linear $R$, i.e. $\sim R^2$. This is due to the fact that both the instability-inducing terms originate from ion irradiation. Both terms are based on the differences in ballistic properties of the alloy constituents. However, one term refers to general alloy sputtering properties (preferential sputtering) while the other refers to a topographic effect (preferential curvature-dependent displacement). Note that their product comes from multiplication of off-diagonal terms of {\textbf A} in Eq. (\ref{matrix}). Off-diagonal elements describe the feedback interactions between the surface roughness and the composition. This brings us to the two conclusions: (i) {\it instabilities can be generated by ion-induced $h(x,y)$-$c_{S}(x,y)$ feedback interactions} and (ii) {\it the strength of the ion-induced feedback interactions is driven by differences of material ballistic properties}. The larger the difference in ballistic properties (sputter/relocation yields and their sensitivity to the surface curvature), the larger is the probability for this effect to occur.

\subsubsection{Instabilities driven by quadratic terms in $q$ (type I)}

The term $a(q)$ can become negative not only due to linear terms in $q$ proportional to $c$ but also due to quadratic terms in $q$ proportional to $b$ (see Eq. (\ref{eigenvalue})). Let us consider the case when $c>0$ and $b<-2\sqrt{c}$. The slope of $a(q)$ is positive close to zero due to the positive linear term $cq$. Only for larger $q$ values the term $\sim -|b| q^2$ counteracts the positive increase leading to $a(q)<0$ (see middle plot (b) of Fig.~\ref{Fig2}). This results in a band of unstable $q$ vectors appearing in a narrow $q$ range with $q>0$. Therefore, close to the instability threshold the wavelength is not diverging. The instability will not occur for any small $R$, but $R$ needs to reach a critical value 
\be
R_{\rm crit,2}\quad {\rm given}\,{\rm by}\quad
b^2=4 c
\label{Rcrit2}
\ee
from Eq. (\ref {bc-coeff}).

The condition $b<0$ requires $S^*_x D >0$ (see Eq. (\ref{bc-coeff})). The terms $S^*_x$ and $D$ describe the preferential curvature-dependent displacement rate and preferential diffusion of the alloy constituents, respectively (see Table \ref{table1}). This represents the following scenario: 

\begin{itemize}
 \item for positive (negative) $S^*_x$ values the preferentially displaced element is depleted from the crests (depressions) and accumulates in the depressions (on the crests) (the term $R S^*_x \frac{\partial^2 U}{\partial x^2}$ in Eq. (\ref{linear-equations}),
 \item such a segregation induces a concentration gradient between crests and depressions, 
 \item a diffusion-driven transport sets in to homogenize the element distribution on the surface (the term $D \nabla^2 \phi$ in Eq. (\ref{linear-equations}),
 \item the element with a larger diffusion coefficient diffuses from the depressions to the crests,
 \item as this element tends to segregate at depressions, preferential diffusion of this element from depressions results in the increase of the surface roughness, i.e. an instability occurs.  
\end{itemize}

Let us assume that close to the instability threshold $R=R_{\rm crit-2}+\delta$ and therefore $(b/2)^2-c=const \times \delta +...>0$. According to Eq. (\ref{solq}), the band width takes the form  $\Delta q =q_0^+-q_0^-= \sqrt{b^2-4 c}$ yielding 

\begin{equation}
\Delta k_x = \sqrt{\Delta q} \sim \delta^{1/4}.
\label{deltak-2}
\end{equation}

The band center can be determined from the condition $ d a(q) / dq =0$. Leaving only linear terms close to the instability results in

\begin{equation}
k_{\rm max} \approx \sqrt{-\frac{1}{6}b_{\rm crit}+\frac{const}{|b_{\rm crit}|} \times \delta}. 
\label{kmax-2}
\end{equation}

The smoothing effect of the curvature-dependent displacement  coefficient $S$ and diffusion $D^*$ separately stabilize the surface and the composition at small and large wave numbers, respectively.  This allows to obtain a narrow band of unstable wave numbers, i.e. the formation of an ordered structure. According to Cross\&Greenside, \cite{Cross:2009} this is characteristic for the so-called Type-I instability. Similar effect has been predicted for ion-erosion of alloy surfaces. \cite{Shenoy:PRL2007} In contrast to the ion-erosion case where the surface atoms are always exposed to ion irradiation, here the interplay of the ion-irradiation and diffusivity is 'damped' by the depositing species (both parameters $R$ and $D$ are divided by $j_{at}$, see Table \ref{table1}).

Note, that the destabilizing effect comes as a linear effect in $R$. This is due to the fact that only one term in the product $S^*_x D$ refers to ion-induced effects (preferential curvature-dependent displacement). Similarly to the instability type II$_b$, the product of terms $S^*_x D$ inducing this type of instability originates from multiplication of off-diagonal terms of {\textbf A} in Eq. (\ref{matrix}) which describe $h(x,y)$-$c_{S}(x,y)$ feedback interactions. The preferentially displaced element must also have a larger diffusivity to keep $S^*_x D > 0$. The above discussion leads to the following conclusion: {\it external preferential ion-induced curvature-dependent processes and material-inherent diffusive processes can couple to induce an instability}. The larger the difference in curvature-dependent ballistic and surface diffusion properties, the larger is the probability for this type of instability to occur.

\subsubsection{Special paths from type-II to type-I instabilities}

By increasing $R$ one can observe a change from type-II instability to a stable behavior and further to the appearance of type-I instability. Such a behavior requires that the total curvature-dependent displacement coefficient has a destabilizing effect on the surface roughness ($S_x<0$) and a stabilizing effect due to synergistic effects of the preferential curvature-dependent displacement rate and the preferential sputtering, $S_x Y^* - S^*_x Y > 0$.

Mathematically, this is due to the fact that by increasing the ion-to-atom arrival ratio $R$, the coefficient $c \sim R \left[S_x + R(S_x Y^* - Y S^*_x) \right]$ (first order in $q$ term) can change the sign from negative (first term of $c$) to positive (second term of $c$). Hence we have the transition: type-II instability$\rightarrow$ stable behavior. With further increase of $R$, the coefficient $b>0$ (second order in $q$ term) might change the sign, which results in type-I  instability. 

The further discussion of these parameters is quite complex and is outside of the scope of the present paper. Our purpose is only to point out here that certain alloy deposition conditions can lead to  type-II$\rightarrow$stable$\rightarrow$type-I transitions when increasing the ion-to-atom arrival ratio $R$.

\subsubsection{Oscillatory instability}

The instability growth exponent $r$ can also change the sign when $f(q)$ changes the sign from positive to negative while $a(q)$ remains positive (see Fig.~\ref{Fig2} (a)). This leads to an increase of the instability amplitude due to  the first term $r\sim - f(q) > 0$. However, the second term of $r \sim \sqrt{f(q)^2 - 4 a(q)}$ (Eq. (\ref{eigenvalue}) remains imaginary, hence we have an oscillating behavior with the increasing amplitude. 
The necessary condition follows from the expression of $f(q)$ in Eq. (\ref{eigenvalue})
\begin{equation}
 e<-2 \sqrt{d}.
\label{e-negative}
\end{equation}
For $e<0$, the total curvature-dependent displacement yields $S_x$ must be negative, thus destabilizing (see Eqs. (\ref{eigenvalue}) and (\ref{bc-coeff})). Like the type-II$_a$ instability, it is also driven by ion-induced surface-roughness processes. 

The oscillatory instability ${\rm Re}(r)>0$ requires exactly the same $q$-range where $f(q)$ becomes negative. The instability occurs when $R$ surpasses a critical value 
\be
 R_{\rm crit,3}\quad {\rm defined}\, {\rm by}\quad e^2=4 d.
\label{Rcrit3}
\ee
Near the critical threshold $R_{\rm crit,3}$ we can expand again $e^2/4-d\approx {\rm const} \times \delta$ and obtain the unstable wave number band width in lowest order
\be
\Delta k^2 =  \Delta q\sim \delta^{1/2}.
\ee

The maximal growth rate is given now approximately by the maximum of $f(q)$ which leads to 
\be
k^2_{\rm max} = q_{\rm max}=-\frac{e}{2} \sim -\frac{e_{crit}}{2} - \frac{S_x}{2B} \delta
\label{type-III-kmax}
\ee
with the value $e_{\rm crit} = \left( D^* + R_{\rm crit,3} S_x \right) / B$
(Eq. (\ref{bc-coeff})). Note that the characteristics of this type of
instability differs from those of the oscillatory type-III instability from
Cross\&Greenside \cite{Cross:2009}. The real part of $r$ is not zero,
therefore the maximum growth rate occurs at a finite wavelength in
Eq. (\ref{type-III-kmax}). At the onset of the instability where $R=R_{{\rm
    crit},3}$ we have ${\rm Re} \,(r)=-f(q)=0$, and the instability is purely
oscillatory. Therefore the type-III of Cross\&Greenside is a special case of
our type I in the sense of finite $q$ values as an onset.

In more detail, additionally to the condition $e<-2 \sqrt{d}$, two more conditions must be satisfied to render $a(q)>0$, namely
\be
c>0;\quad b^2< 4 c.
\label{conde}
\ee
The condition $S_x<0$ means that $c<0$ for small $R$ which contradicts the necessary conditions Eq. (\ref{conde}) for the oscillating instability to occur. If the parameter $c$ would not change its sign to positive with increasing $R$, then the change of sign of $f(q)$ would result only in the change of the instability type, not in the change from stable to unstable situation. We are limiting our discussion only to the transition stable$\rightarrow$ unstable. Therefore we consider that $c<0 \rightarrow c>0$ by increasing $R$. In addition, the change in the sign of $c$ must occur for $R<R_{crit,3}$.

On one hand, after reaching this stabilizing condition, a further increase in $R$ keeps $a(q)>0$ since both linear and quadratic terms proportional to $c$ and $b$, respectively, are positive. On the other hand, it induces the sign change for $f(q)$, thus the occurrence of the instability. Recalling Eq. (\ref{oscilatory-condition}), one can see that the stabilizing effects are all proportional to diffusion  (both  $D$ and  $\sim \sqrt{B} \sim \sqrt {D}$). Therefore, for the transition type-II$_a\rightarrow$stable$\rightarrow$oscillating instability to occur, the diffusivity must be large enough to allow stabilization of the surface while increasing the ion-to-atom arrival ratio $R$ (transition II$_a\rightarrow$stable). Then further increase in $R$ unavoidably leads to the change of sign of $f(q)$ and hence the transition stable$\rightarrow$oscillating instability. Physically, such a control over surface diffusivity can be achieved independently from the ion-induced effects by changing the substrate temperature. 

The imaginary part of the instability growth rate $r$ is given by ${\rm Im}(r)=\sqrt{4 a(q)-f(q)^2}$. It defines  
the oscillation frequency in thickness units. For the growing film this means that for a given lateral position $(x,y)$, the concentration will oscillate with time. Since the film is growing,  this translates into an oscillatory concentration with thickness. Not only a laterally ordered nanocomposite structure forms, but also vertical composition modulations occur with the period $\approx 1/{\rm Im}(r)$. Such an  oscillating instability establishes a self-organized 3D-multilayer structure. This demonstrates that {\it ion-irradiated surfaces during alloy film growth can induce not only lateral but also vertical periodic structures}.

\subsection{Deposition induced instabilities}

Now we investigate the possibilities to reach the instable regions in Fig.~\ref{Fig2} due to dynamic processes induced by deposition only. 
In the absence of ion irradiation, the terms $c$ and $b$ become $c=S^{\rm dep}/f_{\rm dis}$ and $b=(B+S^{\rm dep} D^*-S^{*\rm dep}D)/f_{\rm dis}$ (Eq. (\ref{bc-coeff})). 
Stationary patterns occur for wave numbers in the interval Eq. (\ref{solq}) and for $c<0$ or $c>0, b<-2 \sqrt{c}$. This translates into two types of instability
\ba
&{\rm Type}\, {\rm IIa:}\,  c<0 \leftrightarrow S^{\rm dep}<0\nonumber\\
&{\rm Type}\, {\rm I:}\,  c>0, b<-2 \sqrt{c} \nonumber\\
&\leftrightarrow S^{\rm dep}>0, S^{\rm dep} D^*-S^{*\rm dep}D<-B-2 f_{\rm dis} \sqrt{S^{\rm dep}}.
\end{align}
This shows that for reaching the type-I instability a sufficient preferential deposition $S^{\rm dep}_A-S^{\rm dep}_B<S_c$ is necessary. We see that for reaching the unstable region of the lower right quarter in Fig.~\ref{Fig2}a only type I and type IIa instabilities are possible from the ones identified above.

Recall, that the the upper left quarter of Fig.~ \ref{Fig2} (a) represents the oscillatory instability of type III above. For such oscillatory patterns we have $e=(D^*+ S^{\rm dep})/B$ and $d=1/B>0$ in the absence of ion irradiation $R=0$. $f(q)$ has to change the sign and $a(q)>0$ which translates into the conditions of Eq. (\ref{conde}).  The condition $c>0$ becomes $S^{\rm dep}>0$ and the second condition of Eq. (\ref{conde}) becomes $D^*+S^{\rm dep}<-2 \sqrt{B}$. This is excluded since the left side is positive. We conclude that {\em no oscillatory instability can be reached at perpendicular atom incidence due to curvature-dependent deposition effects.} Similarly, it was found by Guyer and Voorhees (Fig.~4 in Ref. \onlinecite{Guyer:PRB1996}) that the imaginary part of the instability growth rate is connected with a negative, i.e. damping, real part. For this type of instability to occur one needs oblique angle deposition. \cite{Lichter:PRL1986}

Since in the absence of ion irradiation  ($R=0$) no oscillating instability can be induced due to the deposition effects we can limit ourselves only to the deposition induced type-I and type-II instabilities. These are controlled by the terms linear and quadratic in $q$ proportional to $c$ and $b$, respectively, as discussed above. 

Similarly to the ion-induced instability,  the sum of the dynamic deposition effects must be destabilizing to produce the type-II$_a$ instability. This yields that crests are growing faster than depressions. The band width and the fastest growing wave number are (Eqs. (\ref{deltak-1}) and \ref{kmax-1}) 
\be
\Delta k^2     &\approx & -\frac{S^{\rm dep}}{B} \sim j_{\rm at} \nonumber\\
k_{\rm max}^2  &\approx & -\frac{S^{\rm dep}}{2 B} \sim j_{\rm at}.
\label{deposition-type-II}
\ee

The fastest growing wavelength is diverging for small fluxes. The instability growth is the interplay between the deposition dynamics and the diffusive surface relaxation. The strength of deposition effects increases concomitantly with the deposition flux. The surface diffusivity is constant for a given substrate temperature. The maximum wavelength (wave number) decreases (increases) with the deposition flux since for higher deposition rate the instability-causing fluxes can compete with diffusive-relaxation fluxes at smaller scales.

Again, similarly to curvature-dependent ion-induced instabilities, the type-I instability requires  (i) a sufficient level of curvature-dependent preferential deposition $S^{\rm dep}_A-S^{\rm dep}_B<S_c$ and (ii) a coupling between the curvature-dependent preferential deposition and the preferential diffusivity. The curvature-dependent preferential deposition means that one element accumulates preferentially at the crests or at the depressions. This is a result of different dynamic interaction strengths of alloy components with the topographic features and defects or differences in latent mobilities. The element with larger latent mobility is expected to funnel down more easily. The element with stronger dynamic interactions with the surfaces is expected to be 'steered' towards the crests. The following scenario occurs:
\begin{itemize}
 \item for positive (negative) $S^*_x$ values, the element with stronger (weaker) interaction strength at the surface is depleted from the crests (depressions) and accumulated in the depressions (on the crests),
 \item such a segregation induces a concentration gradient between crests and depressions, 
 \item a diffusion-driven transport sets in to homogenize the element distribution on the surface,
 \item the element with a larger diffusion coefficient diffuses from the depressions to the crests,
 \item again, as this element tends to segregate at depressions, preferential diffusion of this element from depressions results in the increase of the surface roughness, i.e. an instability occurs.  
\end{itemize}

For this effect to occur, a critical deposition flux $j_{\rm at, crit}$ must be reached. Near the instability threshold one can assume $j_{\rm at}=j_{\rm at, crit} + \epsilon$ and the band width takes the form
\begin{equation}
\Delta k_x  \sim \epsilon^{1/4}.
\label{deposition-type-I}
\end{equation}
The fastest growing wavelength has an identical expression as the one obtained for ion-induced instabilities (Eq. (\ref{kmax-2}) with $b_{\rm crit}$ defined at $j_{\rm at, crit}$). Similarly to the ion-induced instability case, the smoothing effect of the deposition dynamics and diffusion separately stabilize the surface and the composition at small and large wave numbers, respectively. This allows to obtain a narrow band of unstable wave numbers, i.e. the formation of an ordered structure.

\subsection{Interplay of ion-irradiation and deposition induced instabilities}

Curvature-dependent ion-induced  displacements and deposition processes can act in the same direction (stabilizing or destabilizing) or in opposite directions (one is stabilizing, another destabilizing). If both processes are destabilizing, then all the conclusions made in the section \ref{section-ion-instabilities} are valid. Mathematically this means that the expression for the curvature-dependent instability driving coefficients $S_x$ and $S^*_x$ must contain terms related to ion-induced and deposition-induced effects. Physically it means that instabilities are controlled by both the atomic flux $j_{\rm at}$  and the ion-to-atom ratio $R$. The critical values to induce type-II$_b$, I and III instabilities will also depend on both of these parameters. 

In the case that the curvature-dependent ion-induced and deposition-induced processes act in opposite directions, the situation differs. For a given atomic flux,  a critical $R_{\rm crit}$ will exist for each type of instability, also for type-II$_a$. This is due to the fact that ions (depositing atoms) have to compensate not only the relaxation effects induced by diffusive fluxes, but also induced by depositing atoms (ions). For the type-II$_a$ instability, which occurs at any small $R$ or $j_{\rm at}$ values in the cases considered above, the critical value is defined by the condition $c=0$. For other types of instabilities the mathematical expressions presented in Section  \ref{section-ion-instabilities} are valid again, but one has always to consider that critical ion-to-atom ratio values must be larger. 

Note that the stabilizing or destabilizing character of ion irradiation is controlled by the ion incidence angle $\theta$. This angle can be adjusted independently from the deposition fluxes. If deposition is destabilizing the film surface, and a homogeneous smooth film growth is desired, the ion incidence angle can be adjusted to produce a stabilizing effect. Then above the critical ion flux value, the ion effects compensate deposition-induced instabilities. In the opposite case when a heterogeneous structure is desired, the ion-incidence angle $\theta$ must be adjusted to act in a destabilizing manner. Though the diffusive and deposition-induced fluxes act in a stabilizing manner, a critical ion flux can be reached when an instability occurs. Similarly, if ion irradiation is used to assist the growing film and its effect is destabilizing, an increase in stabilizing the atomic flux $j_{\rm at}$ can result in a decrease of the ion-to-atom ratio $R$ below the critical value. Homogeneous and smooth growth is produced. This leads to the following conclusion: {\em ion irradiation (deposition) can be used to stabilize the surface roughness and composition if destabilization occurs due to deposition (ion irradiation) effects}.

The surface curvature affects not only instabilities, but also the deposition/re-sputtering balance. This balance is controlled by the terms proportional to sputtering and deposition coefficients  in the total curvature-dependent  rate $S_x$ (see Table \ref{table1}). If dynamic deposition effects result in matter accommodation on (depletion from) some topographic features where sputtering is larger (smaller), a decrease (increase) of the total net growth rate $V$ defined by Eq. (\ref{balance}) results. In a similar manner, surface curvature can affect the steady-state surface concentration $c^0_{S}$. The balance of the species is regulated by the deposition/sputtering terms in the preferential rate $S^*_x$. 

If a certain element is preferentially deposited (depleted) on certain topographic features where its sputtering rate is larger (smaller), a net decrease (increase) of this element happens in comparison to the steady-state value $c^0_{S}$ defined by Eq. (\ref{surf-steady}).  Also any local enrichment  of an element results in the total loss due to the sputtering term $-Y^* \phi$ (see Eq. (\ref{linear-equations})). Note that this balance can be regulated by the ion incidence angle $\theta$ which can change and even inverse the behavior of sputtering for different topographic features \cite{Davidovitch:PRB2007}. This brings us to the conclusion that {\em surface topography does not only influence the local redistribution of the deposited matter but also influences the net growth rate and the average steady state surface composition}.

\section{Phase diagram of experimental parameters}

\subsection{Summary on boundary conditions}

In order to discuss the phase diagram of instabilities induced by curvature-dependent deposition and ion irradiation effects we proceed in two steps. First, we discuss the effects of ion irradiation, i.e. $S^{\rm dep}=0$ exclusively. This will provide us with a unique set of three  parameters which will be expressed as combinations of sputtering and relocation yields. Such parameters will give some insights in the internal structure of the instability driving curvature-dependent terms $S^{\rm ion}_{x,y}$ and $S^{\rm *,ion}_{x,y}$. In a second step we introduce the deposition terms.
  
\subsubsection{Invariant measures}

Now we will link all conditions discussed so far to the original parameters in Table \ref{table1} in a transparent way. To this aim we use three positive combinations of relocation and sputtering,
\be
P_1&=&\frac{S^{\rm rel}_A-S^{\rm rel}_B}{Y^0_A-Y^0_B}\nonumber\\
P_2&=&\frac{c^0_{S}S^{\rm rel}_A+(1-c^0_{S})S^{\rm rel}_{B}}{c^0_{S}Y^0_A + (1-c^0_{S})Y^0_B}\nonumber\\
P_3&=&c^0_{S} \frac{S^{\rm rel}_A}{Y^0_A} + (1-c^0_{S})  \frac{S^{\rm rel}_B}{Y^0_B}
\label{p}
\ee
and rewrite the conditions by equivalent ones as follows
\be
S_x^{\rm ion} Y^* - Y S^{*\rm ion}_x \lessgtr 0 
&\leftrightarrow& S^{\rm sp}_x (\theta) \lessgtr- S^{\rm rel}_x (\theta) P_3,\nonumber\\ 
S_x^{*\rm ion}\gtrless 0&\leftrightarrow&
S^{\rm sp}_x (\theta) \gtrless - S^{\rm rel}_x (\theta) P_1, \nonumber\\
S_x^{\rm ion}\gtrless 0&\leftrightarrow&
S^{\rm sp}_x (\theta) \gtrless - S^{\rm rel}_x (\theta) P_2. 
\label{c-negativ-II}
\ee
One can understand $P_1$ ($P_2$) as the ratio of preferential (total) relocation to preferential (total) sputtering. $P_3$ is the sum of species-dependent relocation and sputtering coefficient ratios. The parameters $P_1, P_3$ are controlling parameters of the surface concentration while the parameter $P_2$ is related to the local height.

Interestingly, these parameters are bounded by 
\be
P_1\gtrless P_2\gtrless P_3 \leftrightarrow
{Y^0_A}{S^{\rm rel}_B} \lessgtr  {Y^0_B}{S^{\rm rel}_A}
\label{relYS}
\ee
and no other possibilities. Here we have assumed that the preferential sputtered species A is also preferentially relocated, so that $P_1>0$. 

The decisive condition ${Y^0_A}{S^{\rm rel}_B} \gtrless {Y^0_B}{S^{\rm rel}_A}$ in Eq. (\ref{relYS}) may seem to be only determined by ballistic properties of given materials A and B, i.e. ratios of sputtering and relocation yields. Therefore low level of control from the experimental parameters (ion energy, type, etc...) may be expected. However, one should note that varying the ion energy close to the sputtering threshold, the relocation yield varies slowly in contrary to the sputtering yield. The latter is zero below the threshold, independently of the relocation yield,  and rises to some finite value above it. This allows to influence significantly the ratios ${Y^0_i}/{S^{\rm rel}_i}$ to match the corresponding requirements for instabilities.

With Eq. (\ref{c-negativ-II}) the conditions for the instabilities can now be translated into relations between sputtering and relocation.

\subsubsection{Ranges of instability types}
\paragraph{Type I}

This type of instability requires $b<0$ as the first  necessary condition. The exact one, $b<-2\sqrt{c}$, is ensured by exceeding the critical value $R>R_{\rm crit,2}$ of Eq. (\ref{Rcrit2}). From the definition of $b$ in Eq. (\ref{bc-coeff}) we see that this means $S_x D^*-S_x^* D<B^*Y-B Y^*<0$. The last inequality follows from Table \ref{table1} which gives $BY^*-B^* Y=[D_B(1-c^0_{S}) Y_A+c^0_{S} D_A Y_B]\rho_S \Delta^2 \gamma / k_B T j_{\rm at}>0$. Furthermore, $B^*\gtrless 0$ always if $D\gtrless 0$ which means preferential diffusion $D_A\gtrless D_B$. 
Consequently, the condition $b<0$ translates into 
\be
S_x<0, S_x^*\gtrless 0 \quad {\rm for} \quad D_A\gtrless D_B.
\ee   
The second necessary condition $c>0$ with $S_x<0$ requires $S_x Y^* - Y S^*_x>0$ and $R>R_{\rm crit,1}$. Using Eq. (\ref{c-negativ-II}) this translates into
\ba
&S^{\rm rel}_x (\theta)\lessgtr 0:\nonumber\\& -S^{\rm rel}_x (\theta)P_{2}>S^{\rm sp}_x>-S^{\rm rel}_x (\theta)P_{3}\quad {\rm for}\quad \frac{Y^0_A}{S^{\rm rel}_A}  \lessgtr \frac{Y^0_B}{S^{\rm rel}_B}\nonumber\\
&S^{\rm rel}_x (\theta)\gtrless 0, D_A>D_B:\nonumber\\& -S^{\rm rel}_x (\theta)P_{2}>S^{\rm sp}_x>-S^{\rm rel}_x (\theta)P_{1}\quad {\rm for}\quad \frac{Y^0_A}{S^{\rm rel}_A} \lessgtr \frac{Y^0_B}{S^{\rm rel}_B}.
\nonumber\\&
\label{c-positive-II-ineq}
\end{align} 

It leads us to the conclusion that $S^{\rm sp}_x (\theta)$ and $S^{\rm rel}_x (\theta)$ must have opposite signs. In the case sputtering acts in stabilizing the surface roughness $S^{\rm sp}_x (\theta)>0$, the relocations must act to destabilize it $S^{\rm rel}_x (\theta)<0$, or the other way around. Therefore the instability can occur only within the ion incidence angle range where sputtering and relocation have opposite impact on the system stability.  As mentioned before, the signs of $S^{\rm rel}_x (\theta)$ and $S^{\rm sp}_x (\theta)$ is governed by different properties of ion-induced collisions cascades and thus occur in different angular ranges. \cite{Davidovitch:PRB2007} Therefore there are large angular ranges where these two quantities have opposite signs. \cite{Davidovitch:PRB2007}

Note that relocation yields are usually much larger than sputtering yields, $S^{\rm rel}_i >> S^{\rm sp}_i$, thus $P_1,P_3>>1$. Maximum values of $S^{\rm sp}_x (\theta)$ and $S^{\rm rel}_x (\theta)$ are of the order of unity. \cite{Davidovitch:PRB2007} This means that not only the signs of $S^{\rm rel}_x (\theta)$ and $S^{\rm sp}_x (\theta)$ must be opposite but also $|S^{\rm rel}_x (\theta)|<<|S^{\rm sp}_x (\theta)|$. The condition of the relocation term being close to zero confines the ion incidence angles around $\theta$-values where the relocation term changes its sign, thus its surface stabilization behavior. This brings to the conclusion that the {\em type-I instability is confined to small angular ranges where $S^{\rm rel}_x (\theta) \approx$ 0}.

\paragraph{Type IIa}

From the discussion above follows, that the curvature-dependent displacement-driven pattern formation is possible for $S_x<0$ and  $S_x Y^* - Y S^*_x<0$ for all $R$  or  $S_x Y^* - Y S^*_x>0$ but $R<R_{\rm crit,1}$ via Eq. (\ref{Rcrit1}).
Therefore we have according to Eq. (\ref{c-negativ-II}) the following possibilities
\ba
&S^{\rm rel}_x (\theta)>0, \forall R:
S^{\rm sp}_x<-S^{\rm rel}_x (\theta)P_{\stackrel{2}{3}}\quad {\rm for}\quad \frac{Y^0_A}{S^{\rm rel}_A}  \lessgtr \frac{Y^0_B}{S^{\rm rel}_B}\nonumber\\
&S^{\rm rel}_x (\theta)<0, \forall R:
S^{\rm sp}_x<-S^{\rm rel}_x (\theta)P_{\stackrel{3}{2}}\quad {\rm for}\quad \frac{Y^0_A}{S^{\rm rel}_A} \lessgtr \frac{Y^0_B}{S^{\rm rel}_B}\nonumber\\
&S^{\rm rel}_x (\theta)\lessgtr 0, R<R_{\rm crit,1}:\nonumber\\&   -S^{\rm rel}_x (\theta) P_{3}<S^{\rm sp}_x<-S^{\rm rel}_x (\theta)P_{2}\quad {\rm for}\quad \frac{Y^0_A}{S^{\rm rel}_A} \lessgtr \frac{Y^0_B}{S^{\rm rel}_B}\nonumber
\nonumber\\&
\label{c-negativ-IIa}
\end{align} 
where we abbreviate the equations such that the upper/lower index of $P$ corresponds to the upper/lower inequality. As expected, this type of instability is not related to $P_1$ which contains in its expression the terms related to preferential relocation or sputtering ( see Eq. (\ref{p})).

\paragraph{Type IIb}

Following Eq. (\ref{c-negativ-II}) we have as condition for the preferential curvature-dependent displacement and sputtering, $S_x>0$ and  $S_x Y^* - Y S^*_x<0$. This leads to 
\ba
&S^{\rm rel}_x (\theta)\gtrless 0, R>R_{\rm crit,1}:\nonumber\\& 
-S^{\rm rel}_x (\theta)P_{2}<  S^{\rm sp}_x<-S^{\rm rel}_x (\theta)P_{3}\quad {\rm for}\quad \frac{Y^0_A}{S^{\rm rel}_A}  \lessgtr \frac{Y^0_B}{S^{\rm rel}_B}.
\nonumber\\&
\label{c-negativ-II-ineq}
\end{align} 
Here we abbreviate again the equations such that the upper/lower inequalities correspond to each other. Although the preferential sputtering is necessary for this type of instability (see Section \ref{section-ion-instabilities}), it does not appear explicitly. {\em The instability is governed by differences in ballistic property ratios of each component}.

\paragraph{Type III}

The necessary conditions for this type instability are $e<0$ which means that $S_x<0$ and $c>0$. With the analogous discussion as above, this translates into the condition 
\ba
&S^{\rm rel}_x (\theta)\lessgtr 0:\nonumber\\& -S^{\rm rel}_x (\theta)P_{2}>S^{\rm sp}_x>-S^{\rm rel}_x (\theta)P_{3}\quad {\rm for}\quad \frac{Y^0_A}{S^{\rm rel}_A}  \lessgtr \frac{Y^0_B}{S^{\rm rel}_B}.
\nonumber\\&
\label{c-positive-III-ineq}
\end{align} 

\subsection{Phase diagram of instabilities due to curvature-dependent ion irradiation effects}

\begin{figure*}[h]
\includegraphics[width=13.5cm]{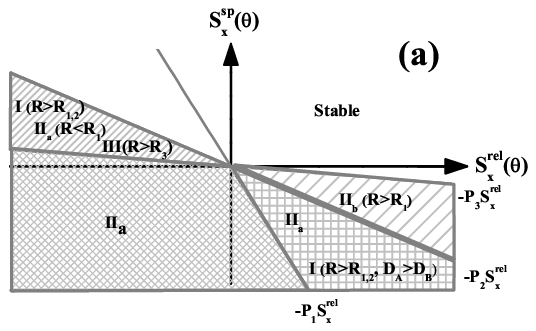}
\includegraphics[width=13.5cm]{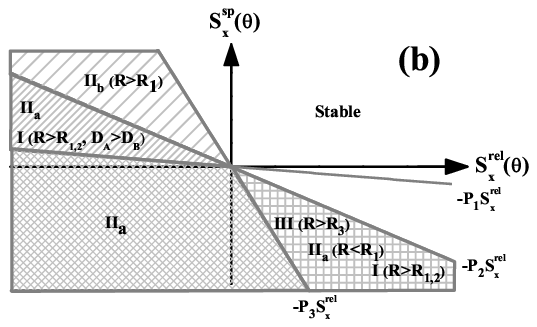}
\caption{\label{diagram1} The phase diagram of different types of instabilities for the relocation to sputtering ratio of species A ${S_A^{\rm rel} \over Y_A^0}> {S_B^{\rm rel}\over Y_B^0}$ (a) and ${S_A^{\rm rel} \over Y_A^0} < {S_B^{\rm rel}\over Y_B^0}$ (b). The critical $R$ parameters are defined by Eqs. (\ref{Rcrit1}), (\ref{Rcrit2}) and (\ref{Rcrit3}). $R_{1,2}$ means $max(R_{\rm crit,1},R_{\rm crit,2})$.}
\end{figure*}

We collect the different identified areas of Eqs. (\ref{c-positive-II-ineq}), (\ref{c-negativ-IIa}), (\ref{c-negativ-II-ineq}),  (\ref{c-positive-III-ineq}) together in Fig.~\ref{diagram1}. The diagram axis are $S^{\rm rel}_x (\theta)$ and $S^{\rm sp}_x (\theta)$. The ion-incidence angle $\theta$ is noted here intentionally to emphasize that it controls the sign of the curvature-dependent relocation and sputtering coefficients. Note again that  $S^{\rm rel}_x (\theta)$ and $S^{\rm sp}_x (\theta)$ change their sign at different ion-incidence angles $\theta$. \cite{Davidovitch:PRB2007} From the discussion above the relative sign changes of sputtering and relocation decides which branch of instability type we have to expect. Physically, different areas of this diagram could be reached by varying the incident ion angle, the energy or the species type. Three critical values of $R_{\rm crit}$ discussed above will determine where the relevant instability types can occur.

Let us first consider the case $\frac{Y^0_A}{S^{\rm rel}_A}  < \frac{Y^0_B}{S^{\rm rel}_B}$. We see that the major range is given by the type IIa instability which is all the area below the $-S_x^ {\rm rel} P_2$ line. Above the $-S_x^ {\rm rel} P_1$ line for $S_x^ {\rm rel}>0$ and above the $-S_x^ {\rm rel} P_3$ line for $S_x^ {\rm rel}<0$ this instability occurs only for a restrictive range of $R$ as indicated in the figure. The type IIb instability occurs only for $S_x^ {\rm rel}>0$ and between the $-S_x^ {\rm rel} P_2$ and $-S_x^ {\rm rel} P_3$ lines. The type I instability can occur for both signs of $S_x^ {\rm rel}$. The type III oscillating instability only for negative $S_x^ {\rm rel}<0$ between the $-S_x^ {\rm rel} P_2$ and $-S_x^ {\rm rel} P_3$ lines. The overlapping regions in the figure are separated by critical values of $R$ via Eqs. (\ref{Rcrit1}), (\ref{Rcrit2}), (\ref{Rcrit3}). 

The analogous figure for the case ${Y_A^0\over S_A^{\rm rel}}>{Y_B^0\over S_B^{\rm rel}}$ is plotted in Fig.~\ref{diagram1} below. One sees that the regions are mirrored at the line $S_x^{\rm rel}=S_x^{\rm sp}$.

\subsection{Phase diagram of instabilities due to combined curvature-dependent ion irradiation and deposition effects}

Now we include the deposition terms by replacing formally 
\be
S^{\rm rel}(\theta) S_A^{\rm rel}&\to& S^{\rm rel}(\theta) S_A^{\rm rel}+{c_0\over c_{S}^0}\cdot {S_A^{\rm dep}\over R}\nonumber\\
S^{\rm rel}(\theta) S_B^{\rm rel}&\to& S^{\rm rel}(\theta) S_B^{\rm rel}+{1-c_0\over 1-c_{S}^0}\cdot {S_B^{\rm dep}\over R}
\ee 
in the definition of the invariant measures Eq. (\ref{p}). This accounts for the deposition-dependent parameters in table \ref{table1}. We have to change therefore
\be
S^{\rm rel}_x{P_i}\to S^{\rm rel}_x P_i+\frac{Q_i}{R}
\label{sh}
\ee
with 
\be
Q_1&=&\frac{{c_0\over c_{S}^0}S^{\rm dep}_A-{1-c_0\over 1-c_{S}^0}S^{\rm dep}_B}{Y^0_A-Y^0_B}\nonumber\\
Q_2&=&\frac{c_0 S^{\rm dep}_A+(1-c_0)S^{\rm dep}_{B}}{c^0_{S}Y^0_A + (1-c^0_{S})Y^0_B}\nonumber\\
Q_3&=&c_0 \frac{S^{\rm dep}_A}{Y^0_A} + (1-c_0)  \frac{S^{\rm dep}_B}{Y^0_B}
\label{Q}
\ee
where analogously to Eq. (\ref{relYS})
\be
Q_1\gtrless Q_2\gtrless Q_3 \leftrightarrow
{Y^0_A}{1-c_0\over 1-c_{S}^0}{S^{\rm dep}_B} \lessgtr  {Y^0_B}{c_0\over c_{S}^0}{S^{\rm dep}_A}.
\label{relQS}
\ee
The invariant measure $Q_3$ represents the ratio between individual curvature-dependent deposition and sputtering weighted with the incoming flux composition. The parameter $Q_1$ is the ratio of curvature-dependent preferential deposition to preferential sputtering. The parameter $Q_2$ represents the ratio of total curvature-dependent deposition and sputtering rates. The parameters $Q_1, Q_3$ are controlling parameters of the surface concentration while the parameter $Q_2$ is related to the local height.  The two possible orderings of $Q_i$ in Eq. (\ref{relQS}) can be seen as preferential curvature-depedent deposition to preferential sputtering ratio of species A.

\begin{figure*}[h]
\includegraphics[width=13.5cm]{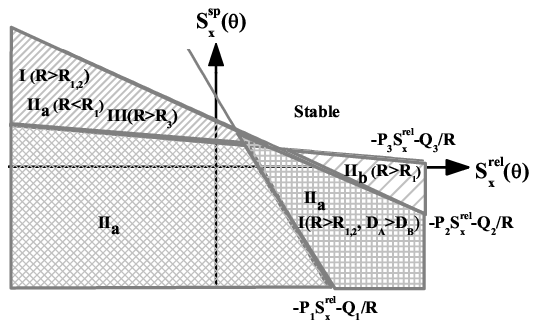}
\caption{\label{diagram2} The phase diagram for preferential relocation to sputtering ratio of species A (as Fig.~\ref{diagram1}a). The deposition effects are included considering the case of preferential curvature-dependent deposition to sputtering ratio of species A. This shifts all axes according to Eq. (\ref{sh}) by the corresponding Q values of Eq. (\ref{Q}).}
\end{figure*}

We see from Eq. (\ref{sh}) that the matter deposition shifts all straight lines in Fig.~\ref{diagram1} by a constant $Q_i/R$. Therefore, the different cases discussed in Fig.~\ref{diagram1} start to overlap and form a new area of phase diagram as seen in Fig.~\ref{diagram2}. This overlap region must contain both conditions from overlapping areas. Choosing first the case of preferential deposition to sputtering ratio of species A, $0>Q_1>Q_2>Q_3$, and the situation of preferential relocation to sputtering ratio of species A in Fig.~\ref{diagram1} (a), the new overlapping region formed by a triangle contains two phases (Fig.~\ref{diagram2})
\be
{\rm type}\, I\, &{\rm with}&\, R>max(R_{\rm crit,1},R_{\rm crit,2}), D_A>D_B\quad {\rm or}\nonumber\\ 
{\rm type}\, IIa\,&{\rm with}&\, R<R_{\rm crit, 1}
\ee
and no oscillatory behavior which is restricted to the left upper region exclusively due to ion-deposition effects. The case of $Q_1>Q_2>Q_3>0$ would shift all lines of Fig.~\ref{diagram2} downwards without any qualitative change of the discussion. The relative shift of the axes are controlled again by the order parameter $R$.

Concerning the case of preferential curvature-dependent deposition to sputtering ratio of species B, $Q_1<Q_2<Q_3$, together with the preferential relocation to sputtering ratio of species A in Fig.~\ref{diagram1} (a), it turns out that the new triangle region has to be an overlap of type II$_a$ and type II$_b$ instability which is impossible. Therefore in this case the triangle region does not specify any instability.

The opposite case of preferential relocation to sputtering ratio of species B in Fig.~\ref{diagram1} (b) just mirrors Fig.~\ref{diagram1} (a) on the axes $S_x^{\rm sp}(\theta)=S_x^{\rm rel}(\theta)$ and all discussions above apply accordingly. 
 
The incident angle of ion irradiation determines the actual values of the curvature-dependent sputtering $S_x^{\rm sp}(\theta)$ and relocation $S_x^{\rm rel}(\theta)$ but not the $P_i$ slopes of the lines. The angle of deposition determines the actual value of the $S_i^{\rm dep}$ parameters and therefore the $Q_i$ offsets. One sees that the actual angle dependencies would determine which of the cases apply. {\em The striking observation is that despite the undetermined and complicated angular dependence of these parameters one can give a phase transition with all possible occurring instabilities}.

\subsection{Relevance to experimental film growth conditions} 

The linear analysis above demonstrates that ion-irradiation during the alloy film growth is expected to play a crucial role on the stability or instability of the growing surface. It is controlled by the ion-to-atom arrival ratio $R$ which can be varied in a broad range suited for numerous experimental conditions. The ion incidence angle $\theta$ determines the sign of $S^{sp} (\theta)$ and $S^{rel} (\theta)$ and thus whether and in which way ions act in destabilizing (stabilizing) the surface roughness and composition.  

Experiments show that energetic ion assistance can act in suppressing the Stranki-Krastanov growth mode of  SiGe alloy films and stabilizing the smooth growth. \cite{Park:JAP1995} An ion-induced stabilizing effect has been also demonstrated for C-Ti nanocomposite films. \cite{Shaha:APL2009,Pei:APL2010,Shaha:MT2011} Moreover, it has been experimentally varified for carbon films that the assisting ion irradiation during ion-beam assisted deposition (IBAD) can cause the formation of the periodic surface roughness patterns. \cite{Zhu:JCP2002,Zhu:JAP2004,Zhu:ASS2006} These experimental findings are in qualitatively agreement with the statement on the importance of ions to control the growing surface morphology.  However, the lack of detailed dependencies of the surface nanopattern structure on the ion-to-atom arrival ratio $R$ does not allow any quantitative comparison with the results of this study. 

Nevertheless, this shows the potential of ion-beam-assisted deposition (IBAD) as a powerful tool to investigate the patterning during the film growth. In contrast to ion-erosion case, it allows to produce nanostructured surfaces whose material is different from that of the substrate. In IBAD the deposition is decoupled from the assisting ion beam. \cite{Nastasi:1996,Smidt:IMR1990} Therefore the control by means of $R$ and $\theta$ is possible. The theory developed in this work shows that these parameters not only induce the formation of a surface pattern, but  can also control their structural properties. Further experimental work is needed to reveal this potential. 

From the theory point of view, it has been demonstrated here that  ion irradiation together with matter deposition can induce phase separation in alloys during ion-assisted film growth. Therefore it has similar effect as the substrate misfit and compositional stresses. \cite{Guyer:PRL1995,Guyer:PRB1996,Guyer:JCG1998,Tersoff:PRL1996,Leonard:PRB1998,Spencer:PRB2001} Similarly to the stress effects, the ion-induced compositional pattern couples to the surface roughness. Compositionally modulated surface roughness patterns occur. Their  behavior close to the instability threshold depends on the regime: driven by ion-induced surface roughness processes (types II$_a$ and III) or by roughness-composition feedback interactions (types I and II$_b$). 

Of particular interest is the possibility to produce patterns with a high degree of ordering.  Our theory predicts that if the conditions for the  type I instability are satisfied, only a  narrow band of unstable wave vectors is produced. This regime is driven by the coupling of preferential curvature-dependent displacement/deposition with preferential diffusivity.  The conditions require that the preferentially displaced material should also be the one with a higher diffusion coefficient. This is not always the case for thermal diffusion. However, at low deposition temperatures ion-irradiation effects on the surface diffusivity can become substantial. \cite{Robinson:JVST1982,Rossnagel:SS1982} This is expected to act on the preferentially displaced element and, therefore, to favor the occurrence of this type of instability. 

On one hand, this opens an alternative opportunity to produce highly compositionally ordered surfaces. On the other hand, such nanopattern is constantly buried by randomly depositing species. The formation of a 3D nanopattern occurs with an ordered heterogeneous structure, hence the control of the resulting nanocomposite morphology. The selected wavelength will be finally determined by non-linear terms stabilizing the initial exponential growth of unstable modes. The tilt of such structures in relation to the film surface will be proportional to the drift terms $\upsilon \cdot \partial h / \partial x$ and $\upsilon^* \cdot \partial h / \partial x$ (see Eq. (\ref{linear-equations})) and thus determined by $R$.


In the Si ion-erosion case, it has been recently demonstrated that sputtering is not needed to induce instabilities. \cite{Madi:PRL2011,Norris:NC2011} While the effect remains to be proven for other material systems, this finding shows that one can work in the regime below or close to the sputtering threshold. Sputtering is to be avoided during the growth due to the loss of material. It limits the ion-to-atom arrival ratio to induce instabilities as complete re-sputtering might occur earlier than any instabilities are induced. Besides, for ions exhibiting energies below the sputtering threshold, film composition is the same as that of the incoming depositing flux $j_{\rm at}$. Following the theory outlined in this work, the ion-induced instabilities are only driven by relocation terms. Assuming a simple angular dependency for the relocation term $\sim {\rm cos}(2 \theta)$ from the literature, \cite{Carter:PRB1996,Davidovitch:PRB2007} a relocation driven pattern growth is expected for ion incidences $\theta > \sim 45^{\circ}$. If alloy growth is desired without formation of any secondary phases, one should grow below the sputtering threshold and at low ion-incidence angles $\theta$.

Note that the above model concerns experimental situations where surface displacements prevail over bulk displacements. \cite{Brice:NIMB1989} It is observed experimentally that ion irradiation also can induce ballistic transport within a 'bulk' of a confined thin sub-surface layer. This results in vertical compositional patterns. \cite{Gerhards:PRB2004,He:PRL2006,Abrasonis:JAP2009,Wu:CPL2004,Chen:APL2010} For comparable fractions of bulk and surface displacements, both mechanisms - ion induced layering and surface instabilities - are expected to act simultaneously, and the produced structure will be the result of the competition between both effects. Our results indicate that vertical layering can also occur only due to surface effects when the oscillating instability occurs.

\section{Conclusions}

Linear instability analysis has been carried out to study the opportunities to induce pattern formation during ion-assisted alloy growth. Curvature-dependent ion-induced relocation and sputtering effects have been considered in the model as well as slope and curvature-dependent dynamic deposition effects. 

It has been demonstrated that during alloy film growth energetic ions (i) can induce phase separation and (ii) drive the system towards stationary or oscillating surface pattern formation. As the surface roughness couples to the composition, this results in compositionally modulated roughness patterns. The control parameter for 'switching' and 'amplifying' such instabilities is identified to be the ion-to-atom arrival ratio $R$. The ion-incidence angle $\theta$ determines whether and in which way the ions have a destabilizing effect on the surface roughness and composition. 

Three different regimes were found close to the threshold driven by surface roughness processes or composition-roughness feedback interactions. The composition-roughness feedback interactions are governed by differences in ballistic, dynamic and thermal properties of the alloy constituents. Two types of stationary instabilities and an oscillatory one have been identified which correspond to the classification presented by Cross and Greenside \cite{Cross:2009}:

\begin{itemize} 
\item Type-I instability which is driven by synergistic interactions of preferential diffusivity and preferential curvature-dependent displacement/deposition rate. It is characterized by a narrow unstable wavelength band starting at finite wavelength whose width $\Delta k$ varies as $\sim (R-R_{\rm crit})^{1/4}$ and whose fastest growing wave number $k_{\rm max}\sim \sqrt{ {\rm const} + (R-R_{crit})}$ near the instability threshold. Such an instability can also occur purely due to deposition dynamics effects and is controlled by the atomic flux $j_{\rm at}$,
\item Type-II instability which is either driven by curvature-dependent surface-roughness processes or by synergistic interactions of preferential sputtering and preferential curvature-dependent displacement/deposition. For the surface-roughness instability, the band of wavelengths and the fastest growing wavelength are characterized both by $\sim \sqrt{R-R_{\rm crit}}$. Such type of an instability can also occur purely due to deposition dynamics effects. If both ion and deposition-driven effects act in inducing instability or one of them is negligible,  both the unstable wavelength band and the fastest growing wavelength are given by $\sim \sqrt{R}$ (or $\sim \sqrt{j_{\rm at}}$). Therefore the wavelength diverges when approaching the instability threshold;
\item
The oscillating or type-III instability occurs due to surface-roughness effects. It is characterized by a wavelength band width $\sim (R-R_{\rm crit})^{1/4}$ and  a finite fastest growing wave number $\sim \sqrt{R-R_{\rm crit}}$ near the threshold. This instability can occur only in the presence of ion irradiation.
\end{itemize}
Oppositely, ion irradiation and deposition can act in a mutually excluding manner resulting in the surface stabilization. 

In addition to causing surface instabilities, surface topography does not only influence the local redistribution of the deposited material but also influences the net growth rate and the average steady state surface composition. This implies that curvature effects must be considered in the deposition balance.

The reason for the fact that ion-induced instabilities can occur via either surface roughness processes or via composition-roughness feedback interactions is that we have considered only alloy systems. This translates into the term $\sim \partial^2 \phi / \partial x^2$ being multiplied by a positive diffusion constant $D^*$ which acts in homogenizing the spatial alloy component distribution. In the case of spinodal decomposition, the coefficient $D^*$ becomes negative, \cite{Cahn:AM1961,Atzmon:JAP1992,Leonard:PRB1997b} and therefore the phase separation occurs spontaneously without any roughness-composition feedback interactions. 

In the case of nucleation and growth, one has to consider nucleation events as finite, not infinitesimal, fluctuations. \cite{Cook:AM1970} Therefore one has to deal with a non-linear theory from the beginning. This can be done by introducing the ion-induced and film growth terms of Eq. (\ref{linear-equations}) into the Cahn-Hilliard-Cook type equations. \cite{Cook:AM1970} Both these cases (spinodal decomposition and nucleation and growth) are more complex and therefore requires more sophisticated approaches. The present work provides a basic framework to address this problem by employing ion-to-atom arrival ratio $R$ as a key control parameter to induce patterning during film growth. We believe that this parameter will play a key role for the phase separating systems.

There is an additional time dependence originating from the ion-induced surface drift which has not been considered in the analysis. This results in a pattern lateral shift as a function of film thickness. While the detailed analysis requires separate theoretical and experimental studies, we note that such drift effects combined with a possibility to rotate the sample during the film growth opens opportunities to sculpt complex 3D structures such as chevrons or helices. Usually the growth of such structures is achieved by using glancing angle deposition. \cite{Robbie:Nature1999,Hawkeye:JVSTA2007} Protection from the degradation of such structures requires a post-growth filling of the spaces between the sculpted nano(micro) objects. The approach outlined in this work would inherently result in an encapsulated nanostructure.

In general, the terms describing other kinetic instabilities during film growth such as Ehrlich-Schwoebel instability have similar mathematical form. \cite{Politi:PR2000} Thus they are also inherently described by the model presented in this manuscript. The physical interpretation is different and has been omitted in the discussion of the present paper. However, such instabilities are relevant in both the contexts - film growth \cite{Politi:PR2000} and ion-induced patterning \cite{Chan:JAP2007}. Therefore, the reiteration of the general findings of this work in these contexts might also provide some new physical insights on how deposition and ion-irradiation effects can synergystically affect the nanopattern formation.

The results of this study present an alternative way to control surface patterns, especially if only a thin nanostructured layer of different material than that of a bulk substrate is needed.  During the film growth the bulk usually remains 'frozen'. Therefore, such a composition pattern formation on the surface during alloy growth translates into an ordered 3D nanocomposite structure. Dependent on the type of stationary or oscillating instability, this can result in laterally or even vertically ordered nanostructures. It should be noted that a self-organized multilayer structure is expected to occur neither due to sequential deposition nor due to sub-surface-driven phase separation \cite{Gerhards:PRB2004,He:PRL2006,Abrasonis:APL2010} but is purely an ion-induced surface effect.

The effects analyzed here are of physical origin. The large energy delivered by ions provides efficient means to compete with other less energetic and material specific processes. We believe that if the ion irradiation is intensive enough (ion-to-atom arrival ratio $R$ is above critical), such effects can be observed in many alloy systems.  This enables a material-independent structural design approach to sculpt the morphology at the nanoscale. As different material systems exhibit different range of properties, this is of possible relevance for a wide range of applications.\\

\acknowledgments

The authors are thankful to Prof. M. R. Bradley from Colorado State University, Fort Collins, United States of America, Prof. W. M\"{o}ller, Dr. K. H. Heinig and Dr. S. Facsko from Helmholtz-Zentrum Dresden-Rossendorf, Dresden, Germany, for enlightening discussions. This work was supported by DFG-CNPq project 444BRA-113/57/0-1 and the DAAD-PPP (BMBF) program. The financial support by the Brazilian Ministry of Science and Technology is acknowledged.

\end{document}